%% file: main.tex
\makeatletter
\@namedef{ver@breakurl.sty}{2099/01/01}
\@namedef{opt@breakurl.sty}{hyphenbreaks}
\makeatother

\documentclass[sn-nature]{sn-jnl}

\def\revcleanbuild{}
\usepackage{fontspec}
\input{preamble}

\usepackage[switch]{lineno}

\makeatletter
\renewcommand{\footnoterule}{%
  \kern-\dimexpr5.5pt+0.2mm\relax
  \ifodd\c@page\hrule\@height0.2mm\@width\textwidth
  \else\hrule\@height0.2mm\@width37mm\fi
  \kern5.5pt}
\makeatother

\renewcommand{\floatpagefraction}{0.6}
\renewcommand{\snRunningTitle}{Complete virtual unwrapping and reading of a rolled Herculaneum papyrus}
\hypersetup{pdftitle={Complete virtual unwrapping and reading of a rolled Herculaneum papyrus}}

\begin{document}

\input{sections/00-frontmatter}

\input{sections/01-technical-strategy}
\input{sections/02-results}
\FloatBarrier
\input{sections/03-discussion}
\FloatBarrier

\bibliographystyle{sn-nature}
\bibliography{references-stages}

\clearpage %
\input{sections/05-methods}
\input{sections/05b-methods-validation}
\input{sections/06-transcription}
\input{sections/05c-methods-reporting}

\input{sections/07-backmatter}

\clearpage
\input{sections/08-extended-data}

\clearpage
\input{sections/09-supplementary}

\end{document}

%% file: preamble.tex
\usepackage{polyglossia}
\setmainlanguage{english}
\setotherlanguage[variant=ancient]{greek}   %

\newfontfamily\headfont{LibertinusSerif-Regular.otf}[
  BoldFont       = LibertinusSerif-Semibold.otf,
  ItalicFont     = LibertinusSerif-Italic.otf,
  BoldItalicFont = LibertinusSerif-SemiboldItalic.otf,
  Letters        = SmallCaps,
  Ligatures      = TeX ]

\newfontfamily\greekfont{OldStandard-Regular.otf}[
  BoldFont    = OldStandard-Bold.otf,
  ItalicFont  = OldStandard-Italic.otf,
  Script      = Greek ]

\usepackage{booktabs}
\usepackage{tabularx}
\usepackage{multirow}
\usepackage{makecell}
\usepackage{rotating}
\usepackage{pdflscape}   %
\usepackage{adjustbox}
\usepackage{array}
\usepackage{ragged2e}    %

\usepackage{graphicx}
\usepackage{xcolor}
\usepackage{float}
\usepackage{caption} %
\usepackage{placeins}     %
\usepackage{needspace}    %
\usepackage{alltt}        %
\usepackage{amsmath}
\usepackage{amssymb}

\makeatletter
\providecommand\headerps@out[1]{}
\makeatother

\definecolor{snaccent}{HTML}{6E1F2E}
\definecolor{snink}{HTML}{1A1A1A}

\hypersetup{
  colorlinks   = true,
  linkcolor    = snaccent,
  citecolor    = snaccent,
  urlcolor     = snaccent,
  filecolor    = snaccent,
  linktoc      = all,
  pdftitle     = {\ifdefined\revproposed Complete virtual unwrapping and reading of a rolled Herculaneum papyrus\else From Hidden Ink to Ancient Thought: Reading Three Still-Closed Herculaneum Scrolls via Virtual Unwrapping\fi},
  pdfauthor    = {Giorgio Angelotti et al.},
  pdfdisplaydoctitle = true,
  unicode      = true }

\usepackage[
  protrusion = true,
  expansion  = false,
  final      = true,
  tracking   = true ]{microtype}
\SetTracking{encoding = {*}, shape = sc}{70}

\makeatletter
\AtBeginDocument{%
  \clubpenalty        = 10000   %
  \widowpenalty       = 10000   %
  \displaywidowpenalty= 10000   %
  \brokenpenalty      = 10000   %
  \interlinepenalty   = 0       %
  \setlength{\emergencystretch}{1.5em}%
}%
\@secpenalty\z@
\makeatother

\makeatletter
\def\sectionfont{\reset@font\headfont\fontsize{15.5bp}{16bp}\bfseries\scshape
  \selectfont\raggedright\color{snaccent}}%
\def\subsectionfont{\reset@font\headfont\fontsize{12bp}{14bp}\bfseries\scshape
  \selectfont\raggedright\color{snink}}%
\def\subsubsectionfont{\reset@font\headfont\fontsize{11bp}{13bp}\scshape
  \selectfont\raggedright\color{snink}}%
\def\bmheadfont{\reset@font\headfont\fontsize{10bp}{12bp}\scshape
  \selectfont\raggedright\color{snink}}%
\def\abstractheadfont{\reset@font\headfont\fontsize{9.5bp}{11bp}\scshape
  \selectfont\titraggedcenter\color{snink}}%
\renewcommand\section{\@startsection{section}{1}{\z@}%
  {-15.5pt \@plus -4pt \@minus -2pt}{8pt}{\sectionfont}}%
\renewcommand\subsection{\@startsection{subsection}{2}{\z@}%
  {-13pt \@plus -4pt \@minus -2pt}{5pt}{\subsectionfont}}%
\renewcommand\subsubsection{\@startsection{subsubsection}{3}{\z@}%
  {-12pt \@plus -4pt \@minus -2pt}{5pt}{\subsubsectionfont}}%
\pretocmd{\section}{\Needspace*{3\baselineskip}}{}%
  {\PackageWarning{preamble}{needspace guard not added to \string\section}}%
\pretocmd{\subsection}{\Needspace*{3\baselineskip}}{}%
  {\PackageWarning{preamble}{needspace guard not added to \string\subsection}}%
\pretocmd{\subsubsection}{\Needspace*{3\baselineskip}}{}%
  {\PackageWarning{preamble}{needspace guard not added to \string\subsubsection}}%
\makeatother

\DeclareCaptionLabelSeparator{snbar}{\nobreakspace\textbar\space}
\makeatletter
\g@addto@macro\captionsenglish{\renewcommand{\figurename}{Fig.}}
\makeatother
\AtBeginDocument{\renewcommand{\figurename}{Fig.}}%

\usepackage{etoolbox}%
\makeatletter
\def\Titlefont{\reset@font\fontsize{19bp}{22bp}\selectfont\titraggedcenter}%
\def\Authorfont{\reset@font\fontsize{10bp}{12bp}\selectfont\boldmath\titraggedcenter}%
\def\addressfont{\reset@font\headfont\fontsize{9.5bp}{12bp}\scshape\selectfont
  \titraggedcenter\color{snink}}%
\newif\if@snaffilstarted \@snaffilstartedfalse
\def\sn@affilsep{\if@snaffilstarted ;\ \else \global\@snaffilstartedtrue \fi}%
\renewcommand{\@@address}[2][]{%
  \g@addto@macro\auaddress{%
     \stepcounter{affn}%
     \xdef\@currentlabel{\theaffn}%
     \jmkLabel{\theaffn}%
     {\sn@affilsep\textsuperscript{#1}#2}}%
}%
\renewcommand{\@@coraddress}[2][]{%
  \g@addto@macro\auaddress{%
     \stepcounter{affn}%
     \xdef\@currentlabel{\theaffn}%
     \jmkLabel{\theaffn}%
     {\sn@affilsep\textsuperscript{#1*}#2}}%
}%
\patchcmd{\@maketitle}{\addressfont\auaddress\par}%
  {\addressfont\auaddress.\par}%
  {}{\PackageWarning{preamble}{affiliation closing period not injected (anchor changed)}}%
\newcommand{\snabstractrule}{\par\addvspace{6pt}%
  \centerline{\color{snaccent}\rule{\textwidth}{0.4pt}}%
  \addvspace{3pt}}%
\patchcmd{\@maketitle}{{\printabstract\par}}{\snabstractrule{\printabstract\par}}%
  {}{\PackageWarning{preamble}{title-page abstract rule not injected (anchor changed)}}%
\patchcmd{\@maketitle}{\removelastskip\vskip20pt\nointerlineskip}%
  {\removelastskip\vskip10pt\nointerlineskip}%
  {}{\PackageWarning{preamble}{title skip not tightened (anchor changed)}}%
\patchcmd{\@maketitle}{\vskip7pt\addressfont}{\vskip4pt\addressfont}%
  {}{\PackageWarning{preamble}{affiliation skip not tightened (anchor changed)}}%
\patchcmd{\@maketitle}{\removelastskip\vskip24pt}{\removelastskip\vskip8pt}%
  {}{\PackageWarning{preamble}{corr-author skip 1 not tightened (anchor changed)}}%
\patchcmd{\@maketitle}{\removelastskip\vskip24pt}{\removelastskip\vskip8pt}%
  {}{\PackageWarning{preamble}{corr-author skip 2 not tightened (anchor changed)}}%
\newif\if@sncorrstarted   \@sncorrstartedfalse
\newif\if@sncontribstarted\@sncontribstartedfalse
\def\email#1{%
  \global\advance\emailcnt by 1\relax
  \if@corauemail
    \if@sncorrstarted \g@addto@macro\corrauthemail{;\ }\fi
    \global\@sncorrstartedtrue
    \g@addto@macro\corrauthemail{\setcounter{footnote}{0}{\color{snaccent}#1}}%
  \else
    \if@sncontribstarted \g@addto@macro\authemail{;\ }\fi
    \global\@sncontribstartedtrue
    \g@addto@macro\authemail{\setcounter{footnote}{0}{\color{snaccent}#1}}%
  \fi}%
\def\abstractfont{\reset@font\fontsize{9bp}{10.5bp}\selectfont
  \leftskip=24pt\rightskip=24pt\parfillskip=0pt plus 1fil}%
\makeatother

\AtBeginEnvironment{sidewaystable}{\setlength{\tabcolsep}{2pt}}

\usepackage{fancyhdr}
\newcommand{\snheadfont}{\normalfont\headfont\fontsize{8bp}{9.5bp}\scshape
  \selectfont\color{snink}}
\newcommand{\snRunningTitle}{\ifdefined\revproposed Complete virtual unwrapping and reading of a rolled Herculaneum papyrus\else Reading Three Still-Closed Herculaneum Scrolls via Virtual Unwrapping\fi}%
\newcommand{\snRunningAuthor}{Angelotti et al.}%
\makeatletter
\fancypagestyle{snrunning}{%
  \fancyhf{}%
  \fancyhead[LE]{\ifnum\value{page}>\@ne \snheadfont\snRunningTitle \fi}%
  \fancyhead[RO]{\ifnum\value{page}>\@ne \snheadfont\snRunningAuthor\fi}%
  \fancyhead[LO,RE]{}%
  \fancyfoot[C]{\ifnum\value{page}>\@ne \snheadfont\thepage\fi}%
  \renewcommand{\headrulewidth}{0.4pt}%
  \renewcommand{\headrule}{\ifnum\value{page}>\@ne
    {\color{snaccent}\hrule\@height\headrulewidth\@width\headwidth
     \vskip-\headrulewidth}\fi}%
}%
\AtBeginDocument{%
  \pagestyle{snrunning}%
  \g@addto@macro\maketitle{\pagestyle{snrunning}}%
}%
\makeatother

\newenvironment{transcription}
  {\par\addvspace{0.6\baselineskip}\begingroup
   \linespread{1.0}\small\setlength{\parindent}{0pt}%
   \begin{alltt}\greekfont}%
  {\end{alltt}\endgroup\addvspace{0.6\baselineskip}}

\newlength{\transtoplead}
\newsavebox{\transboxL}%
\newsavebox{\transboxR}%
\newlength{\transrulew}
\newlength{\transmaxht}%
\newlength{\transmaxdp}%
\definecolor{transrule}{named}{black}%
\newlength{\transwL}
\newlength{\transwR}
\newenvironment{transpair}
  {\begin{lrbox}{\transboxL}%
     \begin{minipage}[t]{\transwL}%
       \ignorespaces}
  {    \end{minipage}%
   \end{lrbox}%
   \setlength{\transmaxht}{\ht\transboxL}%
   \ifdim\ht\transboxR>\transmaxht \setlength{\transmaxht}{\ht\transboxR}\fi
   \setlength{\transmaxdp}{\dp\transboxL}%
   \ifdim\dp\transboxR>\transmaxdp \setlength{\transmaxdp}{\dp\transboxR}\fi
   \par\nobreak\vskip\medskipamount\noindent
   \usebox{\transboxL}%
   \hfill
   \ifdim\dp\transboxR>0.5pt
     {\color{black!25}\rule[-\transmaxdp]{\transrulew}{\dimexpr\transmaxht+\transmaxdp\relax}}%
     \hfill
   \fi
   \usebox{\transboxR}%
   \par\medskip}
\newcommand{\transversion}{%
       \end{minipage}%
     \end{lrbox}%
   \begin{lrbox}{\transboxR}%
     \begin{minipage}[t]{\transwR}%
       \vspace*{\transtoplead}\raggedright\small\ignorespaces}

\ifdefined\refereebuild
  \usepackage{lineno}
  \usepackage{setspace}
  \AtBeginDocument{\linenumbers\doublespacing}
\fi

\definecolor{revred}{HTML}{C00000}
\newif\ifrevcolor \revcolortrue
\ifdefined\revcleanbuild \revcolorfalse \fi

\definecolor{rcover}{HTML}{D00000}   %
\definecolor{rcwarn}{HTML}{E07000}   %
\definecolor{rclang}{HTML}{E0409A}   %
\definecolor{rccite}{HTML}{7B3FB5}   %
\definecolor{rcslip}{HTML}{8B5A2B}   %
\definecolor{rcorder}{HTML}{008B8B}  %
\definecolor{rcgood}{HTML}{1E8E3E}   %
\definecolor{rcneutral}{HTML}{1F5FBF}%

\ifdefined\revclassified \definecolor{revred}{HTML}{1F5FBF}\fi
\usepackage{eso-pic}

\definecolor{rcnew}{HTML}{008B8B}

\usepackage[normalem]{ulem}
\definecolor{pctit}{HTML}{1F5FBF}
\definecolor{pccred}{HTML}{1E8E3E}
\definecolor{pcsoft}{HTML}{B08800}
\definecolor{pcfig}{HTML}{6B3A16}
\definecolor{pclim}{HTML}{8E1B4A}
\definecolor{pcfix}{HTML}{3B2E9E}

\ifdefined\revproposed
  \renewcommand{\floatpagefraction}{0.6}
\fi
\usepackage{xltabular}

%% file: sections/00-frontmatter.tex
\title{Complete virtual unwrapping and reading of a rolled Herculaneum papyrus}

\author*[1]{\fnm{Giorgio} \sur{Angelotti}}\email{giorgio@scrollprize.org}
\author[1,2]{\fnm{Stephen} \sur{Parsons}}
\author[1,3]{\fnm{Federica} \sur{Nicolardi}}
\author[1]{\fnm{Youssef} \sur{Nader}}
\author[1]{\fnm{Sean} \sur{Johnson}}
\author[1]{\fnm{David} \sur{Josey}}
\author[1,4]{\fnm{Paul} \sur{Henderson}}
\author[1]{\fnm{Hendrik} \sur{Schilling}}
\author[1]{\fnm{Johannes} \sur{Rudolph}}
\author[1]{\fnm{Forrest} \sur{McDonald}}
\author[1]{\fnm{Elian Rafael} \sur{Dal Prá}}
\author[5]{\fnm{Paul} \sur{Tafforeau}}
\author[5]{\fnm{Alessandro} \sur{Mirone}}
\author[2]{\fnm{C. Seth} \sur{Parker}}
\author[1]{\fnm{Jan Paul} \sur{Posma}}
\author[1]{\fnm{Benjamin} \sur{Kyles}}
\author[1,3]{\fnm{Claudio} \sur{Vergara}}
\author[1,3]{\fnm{Alessia} \sur{Lavorante}}
\author[1,3]{\fnm{Rossella} \sur{Villa}}
\author[1,3]{\fnm{Maria Chiara} \sur{Robustelli}}
\author[1,3]{\fnm{Marzia} \sur{D’Angelo}}
\author[1,6]{\fnm{Gianluca} \sur{Del Mastro}}
\author[1,7]{\fnm{Michael} \sur{McOsker}}
\author[1,8]{\fnm{Kilian} \sur{Fleischer}}
\author[1,2]{\fnm{Christy} \sur{Chapman}}
\author[1]{\fnm{Nat} \sur{Friedman}}
\author[1,2]{\fnm{W.~Brent} \sur{Seales}}

\affil[1]{Vesuvius Challenge, San Francisco, CA, USA}
\affil[2]{EduceLab, University of Kentucky, Lexington, KY, USA}
\affil[3]{Università degli Studi di Napoli Federico II, Napoli, Italy}
\affil[4]{University of Glasgow, Glasgow, UK}
\affil[5]{ESRF, Grenoble, France}
\affil[6]{Università degli Studi della Campania Luigi Vanvitelli, S. Maria Capua Vetere, Italy}
\affil[7]{University College London, London, UK}
\affil[8]{Universität Tübingen, Tübingen, Germany}

\abstract{The carbonized papyrus scrolls from Herculaneum preserve the only large-scale library to survive from classical antiquity. Many scrolls have remained unread because physical unrolling risks irreversible damage. Non-invasive imaging offers a route to their texts, yet previous work on sealed Herculaneum scrolls has recovered only localized readings. Here, by combining high-resolution X-ray computed microtomography (µCT) data acquired on the BM18 beamline at the ESRF with machine-learning-assisted virtual unwrapping, we study three still-closed scrolls. For \textit{PHerc.}~1667, a remnant of earlier attempts at physical unrolling, we achieve the complete unwrapping and reading of the preserved writing surface, making it the first Herculaneum papyrus read completely by virtual unwrapping. The text was transcribed and cross-verified by nine papyrologists. In \textit{PHerc.~Paris.}~4, the source of the Vesuvius Challenge 2023 Grand Prize text, the high-resolution scan makes ink directly visible in the tomographic volume, allowing 3D ink segmentation and comparison of recovered text with visible deposits. In \textit{PHerc.}~139, we recover title and author-attribution evidence identifying the scroll as Philodemus, \emph{On Gods}, Book~8. These results establish a basis for the ongoing, systematic recovery and scholarly study of texts preserved within the still-closed Herculaneum library.}
\maketitle

%% file: sections/01-technical-strategy.tex
The Herculaneum papyri present one of the longest-standing readability challenges in the study of the ancient world. Discovered in the eighteenth century in the Villa of the Papyri, they make up the only surviving large-scale written library from classical antiquity\cite{ref1}. Carbonized by the eruption of Mount Vesuvius in AD 79, the rolls survived in a form that made reading and preservation fundamentally opposed: physical opening could reveal text, yet also irreversibly damage the material. Earlier campaigns of mechanical opening revealed extraordinary texts, especially Epicurean works, but also demonstrated the destructive cost of direct intervention. Hundreds of rolls and multi-layer roll fragments therefore remain sealed, with their contents preserved yet inaccessible\cite{ref1}.

Virtual unwrapping offers a non-invasive, computational approach to revealing these texts\cite{ref15,ref16,ref17,ref18,ref19,ref20,ref21,recNeuber2012,scpNew4,scpNew5}. The recovery of two columns of Hebrew text from the severely damaged En-Gedi scroll established a general framework for this approach\cite{ref15}. The En-Gedi framework operates entirely on data from X-ray computed microtomography (µCT), a form of computed tomography (CT), and combines recovery of the writing-surface geometry, rendering of the ink, digital flattening and merging of individual pieces into a coherent, readable image.

The potential impact of virtual unwrapping at Herculaneum is enormous in quantitative terms\cite{fleischer2022}. Of the 659 inventory numbers in the collection referring to unopened material, at least 388 may be sufficiently well preserved to contain substantial continuous texts, including 179 potentially complete or nearly complete scrolls\cite{nicolardi2024}. Assuming an average length of 9–12 metres per scroll\cite{cavallo1983}, these 179 scrolls alone could represent approximately 1.6–2.15 kilometres of papyrus, containing an estimated 21,000–36,000 columns of ancient writing. The significance is all the greater because the texts recovered and identified so far, both from physically unwrapped papyri and through partial attempts at virtual unwrapping, exclusively include works that have no parallel in any other source and were not transmitted through the medieval manuscript tradition. These — and, very likely, those still to be read —  are not simply new copies of known works, but books preserving otherwise lost texts.

Extending the En-Gedi framework to the Herculaneum scrolls requires advances in both imaging and computation, with machine learning integrated into surface localization and ink detection. The extreme physical forces of the Vesuvius eruption resulted in pronounced bends, folds, breaks, and effacement of each scroll’s papyrus structure, all of which pose significant challenges for the geometric modeling and flattening steps of virtual unwrapping. Additionally, the carbonized papyrus substrate on which the writing appears is radiographically similar to the carbon-based inks predominantly used across the Herculaneum collection. This leads to weak or even indistinguishable contrast between the ink strokes and the evidence of papyrus in the tomographic volume. Making matters of weak contrast even worse are the variations in ink recipes, ink contaminants, morphology of the ink deposits, and the state of preservation within and among Herculaneum scrolls. This situation means that writing may be expressed throughout the tomographic volume as a varying combination of faint metallic contrast, surface shape relief, or subtle textural signal rather than by a single, uniform signature.

Earlier studies explored several sources of signal in Herculaneum material, from phase-contrast imaging of rolled papyri to trace metallic components in some inks and fine surface deformations associated with ink on already-opened fragments\cite{ref2,ref3,ref4,ref6,scpNew6}. While these and other studies offered incremental progress toward reading the still-rolled papyri, a complete virtual unwrapping and reading remained out of reach.

We address these challenges through recent advances of the Vesuvius Challenge, an open-science innovation contest launched in 2023. The Vesuvius Challenge brought together computer science researchers, academic subject matter experts, and an online distributed community of contributors, making the challenge public through a deliberately hybrid model of competition and collaboration\cite{scpNew9}. While prior work had established that text could be recovered from unopened or unexposed Herculaneum material\cite{ref16,scpNew7}, the Vesuvius Challenge progressively scaled up the quality and quantity of readable text\cite{ref7,ref8,ref9,ref10}. The First Letters Prize (2023) and the ensuing Grand Prize (2023) revealed multiple columns of readable text from \textit{PHerc.~Paris.}~4\cite{ref10}, while Vesuvius Challenge-sponsored Kaggle competitions expanded the methodological base for addressing the continuing challenges of reliable ink detection\cite{vesuvius-challenge-ink-detection} and accurate surface localization\cite{vesuvius-challenge-surface-detection}. Other Vesuvius Challenge milestones have included non-invasive recovery of the title of \textit{PHerc.}~172 (Bodleian Library, Oxford)\cite{ref13} and semi-automated digital unwrapping of the lower portion of that scroll\cite{ref14}.

Here we substantially extend the En-Gedi framework, combining high-resolution synchrotron µCT with semi-automated surface recovery and machine learning aided ink detection to recover text from three still-rolled Herculaneum scrolls. First, we report the complete virtual unwrapping of the preserved writing surface of \textit{PHerc.}~1667 (National Library of Naples “Vittorio Emanuele III”) along with a complete papyrological reading of the revealed text. Next, with \textit{PHerc.~Paris.}~4 (Académie des Inscriptions et Belles Lettres, Institut de France) we demonstrate dramatic improvements to the resolution and readability of the recovered text in comparison to the previously released Vesuvius Challenge Grand Prize region\cite{ref10}. Further, the high-resolution scan makes the ink more directly visible to the human eye in the tomographic slices, enabling comparison of the model predictions with visible ink deposits. Finally, in \textit{PHerc.}~139 (National Library of Naples “Vittorio Emanuele III”), we recover the title and author identifying the text as Philodemus, \emph{On Gods}, Book~8. Together, these results push the virtual unwrapping of the Herculaneum scrolls beyond proof-of-concept demonstrations to a scalable research framework toward recovering the unopened library.

\raggedbottom
\section*{Technical strategy}

\begin{figure}[H]\centering
  \makebox[\linewidth][c]{\begin{minipage}{183mm}\centering
  \includegraphics[width=150mm]{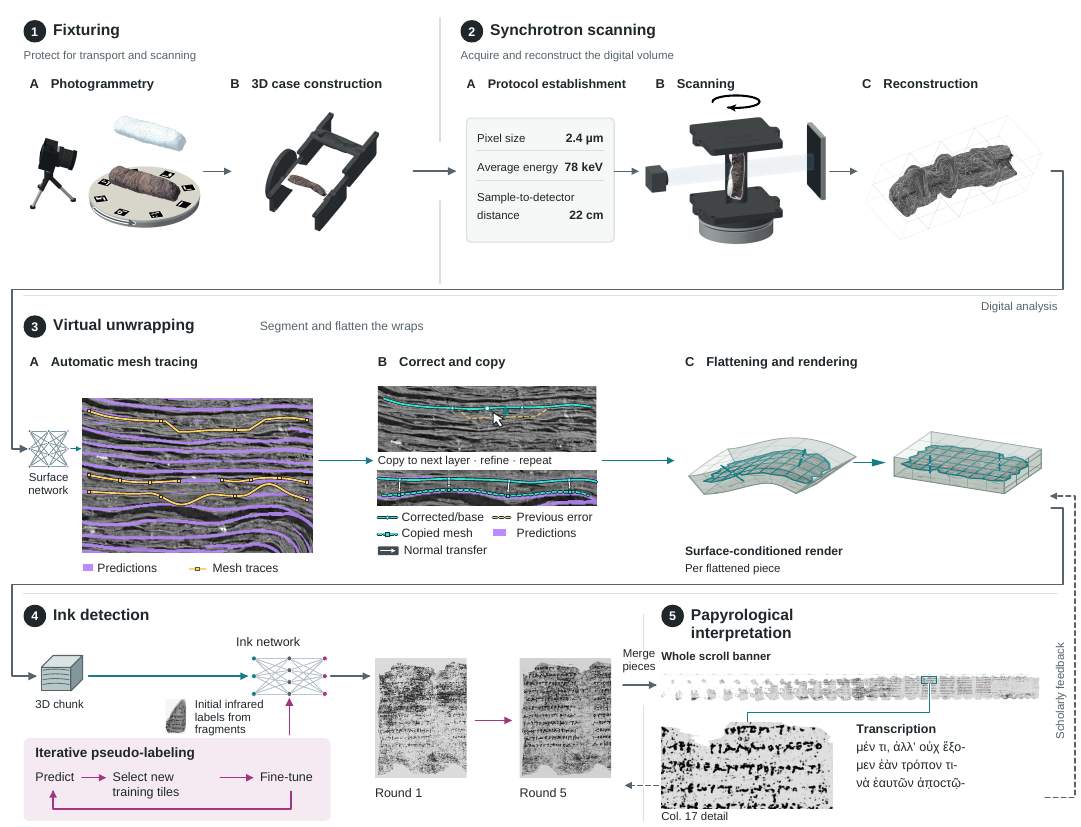}
  \caption{\textbf{Five-stage workflow for virtual unwrapping and reading of \textit{PHerc.}~1667.}
\textbf{1, Fixturing.} Photogrammetry provides an exterior mesh for fitted nylon case construction; protective polytetrafluoroethylene (PTFE) sheet separates the scroll from the case.
\textbf{2, Synchrotron scanning.} Protocol establishment, scanning of the cased scroll and reconstruction produce a digital CT volume. The case is shown in cutaway.
\textbf{3, Virtual unwrapping.} Surface-network predictions guide automatic mesh tracing, followed by correction, copying to adjacent layers, flattening and surface-conditioned rendering of individual pieces. Representative CT and surface-prediction crops carry explanatory mesh overlays; copying and flattening are schematic.
\textbf{4, Ink detection.} Local 3D chunks supply the ink network, initially trained using registered labels from infrared images of fragments. Predictions from selected tiles supply labels for subsequent fine-tuning. Round~1 and Round~5 show predictions on the same held-out column from matched models trained using the corresponding label sets. Flattened, ink-enhanced pieces are merged into a whole-scroll banner.
\textbf{5, Papyrological interpretation.} A banner locator links to a detail of column~17, alongside its transcription excerpt (lines~2--4). Scholars review the image evidence and retain uncertainty in transcription. Dashed feedback paths indicate further review of geometry, rendering or ink enhancement.}
  \label{fig:pipeline1667}
  \end{minipage}}
\end{figure}

We organize the workflow into five stages: fixturing, synchrotron scanning, virtual unwrapping, ink detection and papyrological interpretation (Fig.~\ref{fig:pipeline1667}). The reconstructed CT volume contains the structure of complex carbonized papyrus surfaces on which there is writing. These papyrus surfaces are virtually unwrapped through precise localization (“segmentation”) and structured mesh representations that support flattening and surface-conditioned rendering, which produces a digitally flattened 3D image of the papyrus surface with its immediate surroundings. Machine learning predicts ink from surface-aligned renderings and, for \textit{PHerc.~Paris.}~4, directly from the 3D CT volume before projection onto the flattened surface.

\AddToHookNext{shipout/after}{\flushbottom}

\subsection*{Stage~1: Fixturing}

Photogrammetry uses overlapping photographs to enable a 3D reconstruction of the scroll exterior, from which a fitted nylon transport and scan case is fabricated (Fig.~\ref{fig:pipeline1667}, stage~1A,B; Extended Data Fig.~\ref{fig:ed1}a--c). Under conservator supervision, the scroll is placed in a lined case, wrapped in protective material and secured for mounting in the CT scanner (\hyperref[sec:methods]{Methods}).

\subsection*{Stage~2: Synchrotron scanning}

Prior work indicated that finely detailed spatial resolution (voxel sizes in the single-digit micron range) improves the ability to capture evidence of ink\cite{ref16}. But there are other factors that also play a role in acquiring tomographic scans with enough spatial resolution to reveal ink evidence. For example, variation in ink recipes and weak absorption contrast create a scanning protocol challenge. In addition, the incident energy must be significant enough to penetrate the scroll, but not so energetic that absorption variations are lost. Unwanted X-ray diffusion can occur because the scrolls themselves have regions of densely packed carbonized fibers riddled with micro-structures.

The scan protocol we use balances the need to separate papyrus layers and improve ink resolution within reasonable scan times (Fig.~\ref{fig:pipeline1667}, stage~2A--C). This protocol is the result of an extensive series of imaging tests to calibrate the energy range, the effective resolution, the scanning geometry, and the level of propagation-based phase effects. The outcome of these tests is presented in Extended Data Fig.~\ref{fig:ed2}, while the final acquisition parameters are outlined in \hyperref[sec:methods]{Methods} and \hyperref[tab:ed1]{Extended Data Table~\ref*{tab:ed1}}. Subsequent stages operate on the reconstructed volume and its derived representations.

\subsection*{Stage~3: Virtual Unwrapping}

\begin{figure}[!htb]\centering
  \makebox[\linewidth][c]{\begin{minipage}{160mm}\centering
  \includegraphics[width=\linewidth]{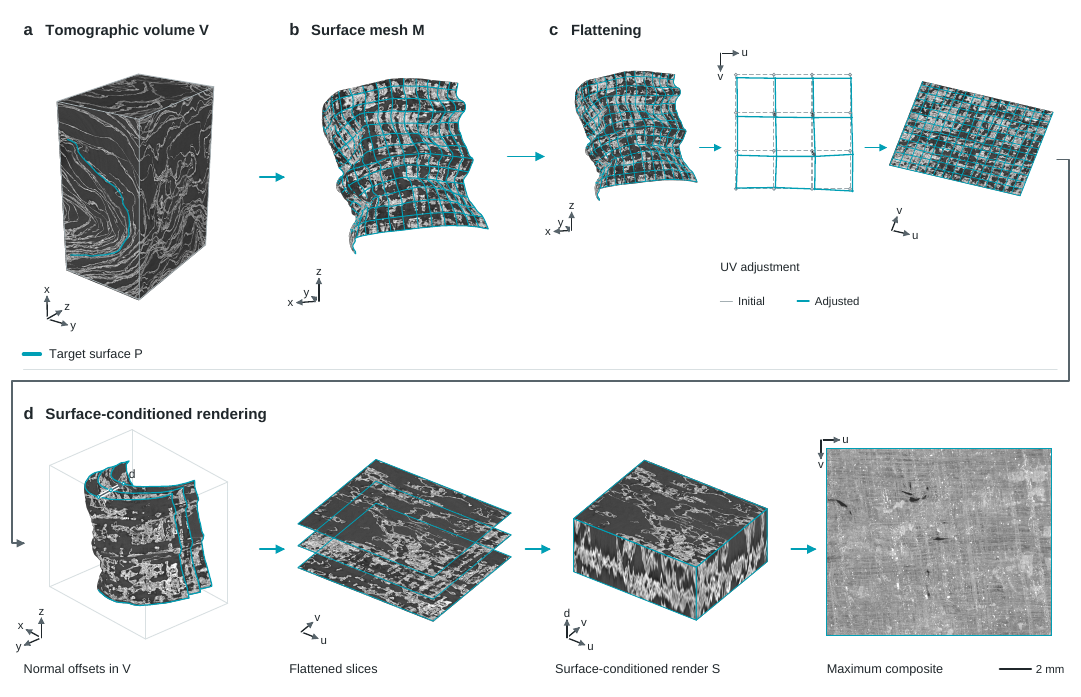}
  \caption{\textbf{Surface geometry and rendering for virtual unwrapping of \textit{PHerc.}~1667.} \textbf{a}, Tomographic volume $V$ with the target surface $P$ indicated in teal. \textbf{b}, A mesh $M$ represents the target surface. \textbf{c}, Flattening maps the surface to coordinates $(u,v)$; the schematic grid illustrates adjustment of the initial coordinates. \textbf{d}, Sampling the CT volume at offsets $d$ along the local surface normal produces aligned flattened slices and the surface-conditioned render $S$. The rightmost panel shows a maximum composite of flattened slices over the depth axis. Scale bar, 2~mm (rightmost panel).}
  \label{fig:surface1667}
  \end{minipage}}
\end{figure}

Virtual unwrapping identifies evidence of the physical papyrus surface of interest ($P$) captured within the tomographic volume ($V$) and reorganizes its neighborhood into a surface-conditioned render ($S$), which is a 3D CT volume centered on the writing surface in flattened coordinates. To create $S$, a learned recto-surface segmentation model first proposes the subset of voxels in the volume $V$ that likely contain the relevant papyrus sheet surface $P$. This selected surface is then modeled as an explicit quadrilateral mesh ($M$). The mesh is grown automatically using orientation evidence derived from the proposed voxel set, manually corrected in those places where it does not adhere to the target layer $P$, and then locally refined to improve agreement with $P$ while preserving surface continuity (Fig.~\ref{fig:pipeline1667}, stage~3A,B). A high-quality mesh can also be offset along its normals to a neighboring surface prediction and refined there. Finally, the mesh $M$ is reparametrized to 2D while constraining for low in-plane metric distortion, minimizing local stretch and shear, to produce a bidirectional map between the 3D surface and its 2D counterpart. Each point in the flattened map retains corresponding 3D coordinates in the µCT volume, allowing CT values to be sampled along the local surface normal (Fig.~\ref{fig:pipeline1667}, stage~3C; Fig.~\ref{fig:surface1667}a--d).

In the surface-conditioned render $S(u,v,d)$, $u,v$ follow the flattened mesh and $d$ measures the offset along its local normal. This warped 3D CT volume forms a stack of aligned 2D slices, each retaining its correspondence to the original scan. Individual slices or projections provide images for inspection (Fig.~\ref{fig:surface1667}d).

From 2D views of the surface-conditioned renders $S$, users may inspect the underlying papyrus layer for evidence of ink, assess the local geometric reliability of the segmentation, and distinguish surface-placement errors from those imposed by ink-enhancement or ink-interpretation processes\cite{ref15,ref16,ref17,ref18,ref19,ref20,ref21,ref24}.

\subsection*{Stage~4: Ink detection}

There is wide variability in how evidence of ink appears across multiple closed scroll regions. In many cases ink is not reliably separable as a 3D object by direct volumetric segmentation within the tomographic volume. When it is, as in the case of \textit{PHerc.~Paris.}~4, our approach to ink detection can leverage the direct observability of ink traces. This scenario admits a direct 3D ink segmentation approach, allowing the identified ink in 3D to project onto virtually unwrapped (digitally flattened) surfaces (\hyperref[sec:methods]{Methods}). For \textit{PHerc.~Paris.}~4, inference is performed across the full reconstructed volume $V$ in parallel with geometric tracing, and the resulting ink volume is subsequently sampled using the flattened surfaces.

More subtle cases require the ink evidence to be extracted through an enhancement process designed with supervised artificial neural models. The models take 3D chunks of $S$ as input and predict 2D ink maps in the flattened $(u,v)$ plane, where the supervision is defined. The models train over two related but distinct label types: direct-visibility labels and pseudo-labels. Direct-visibility labels are used when the presence of ink can be discerned using infrared photographs of opened Herculaneum fragments that are registered to the tomographic scans of those same fragments. Pseudo-labels, on the other hand, are applied to 2D images derived from the surface-conditioned renders $S$ of the closed scrolls, where ink presence is conservatively identified through the manual visual inspection of the surface renders and the output of previously trained ink models. Selected predictions supply pseudo-labels for successive fine-tuning rounds, progressively refining ink detection (Fig.~\ref{fig:pipeline1667}, stage~4; \hyperref[sec:methods]{Methods}). The resulting ink-probability map enhances contrast for papyrological inspection. It is important to note, however, that our models are trained to detect \emph{localized ink presence only} within microscopically small regions; they are \emph{not trained} on character identities, words, transcriptions, OCR or HTR targets, or lexical labels. Together, fragment-derived supervision and restricted spatial context provide safeguards against hallucinated letterforms, directing the model toward local tomographic evidence rather than the completion of plausible words.

These two approaches to ink detection allow model predictions to be compared with ink deposits directly visible in the tomographic volume. The accuracy of ink enhancement can be tested by investigating how well the images of ink it produces align with the physically visible ink deposits rather than with artifacts introduced by rendering, surface placement, or the enhancement model.

\subsection*{Stage~5: Papyrological interpretation}

The separately flattened and ink-enhanced pieces are merged into a single scroll view for papyrological reading (Fig.~\ref{fig:pipeline1667}, stages~4--5). Papyrologists review the recovered text by inspecting the flattened renderings, ink-enhanced outputs and geometric context of the sampled surface. Accepted readings require agreement of the sampled surface with the papyrus layer, consistency of letterforms across repeated renderings or inference passes, and independent endorsement or consensus. The transcription retains underdots for uncertain letters and square brackets for restorations across physically lost substrate. Scholarly review can prompt reinspection and refinement of surface geometry, rendering or ink enhancement, following the dashed feedback paths from stage~5 to stages 3 and 4 (Fig.~\ref{fig:pipeline1667}, stages~3--5).

%% file: sections/02-results.tex
\raggedbottom
\section*{Results}

The results are organized around three questions. First, can the preserved portion of a still-rolled Herculaneum papyrus be completely unwrapped and read? The presented work on \textit{PHerc.}~1667 addresses this question. Second, can ink recovery be checked against ink that is directly visible in the CT volume? The recent scan of \textit{PHerc.~Paris.}~4 provides this imaging-validation case. Third, can virtual unwrapping recover evidence to identify a work preserved within a sealed scroll? \textit{PHerc.}~139 provides this title-identification case.

\subsection*{\textit{PHerc.}~1667: complete unwrapping}

\begin{figure}[H]\centering
  \includegraphics[width=0.85\linewidth]{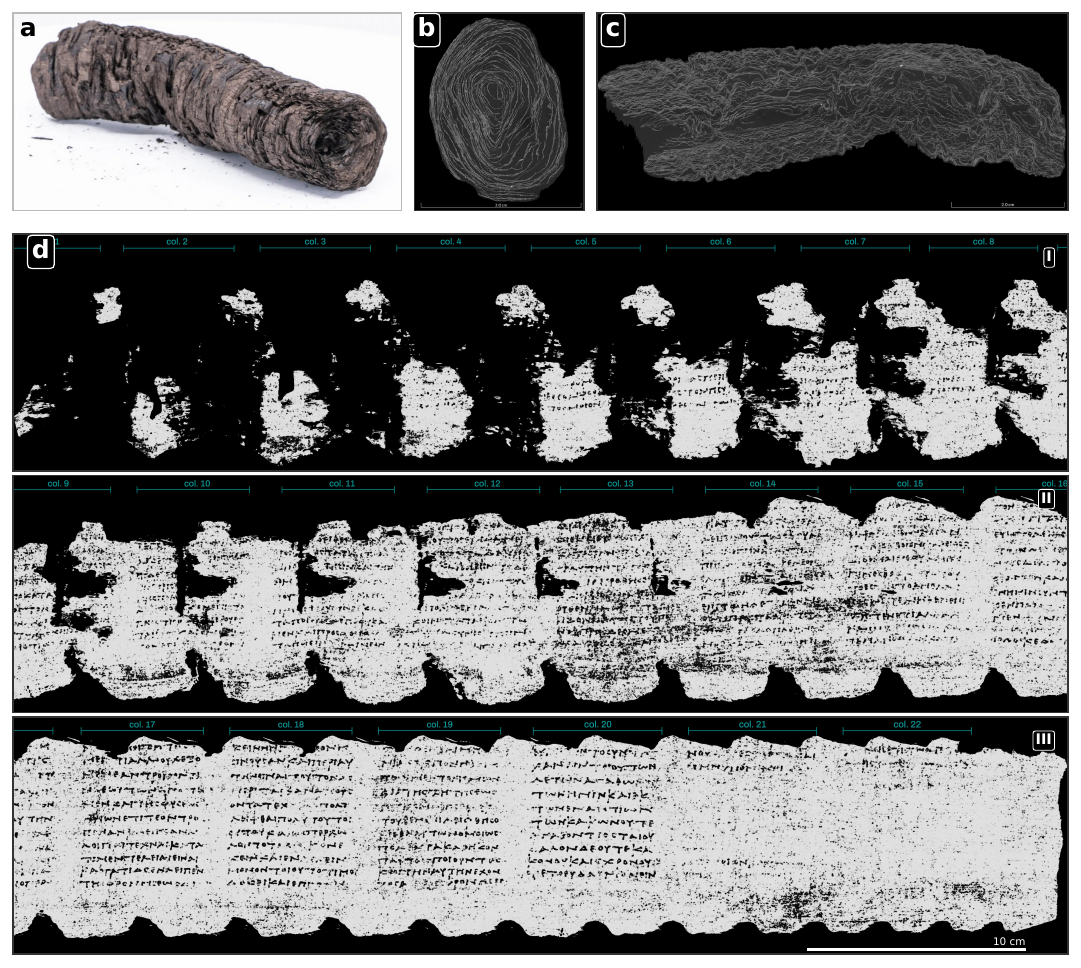}
  \caption{\textbf{Scroll-scale virtual unwrapping of \textit{PHerc.}~1667.} a, Photograph of the sealed Herculaneum roll \textit{PHerc.}~1667. b, Transverse tomographic section (isotropic voxel size 2.4~µm) showing the nested internal wrap structure of the carbonized papyrus. c, Longitudinal tomographic slice showing the layered structure along the scroll axis. d, 2D ink-enhanced image derived from the surface-conditioned render of the reconstructed writing surface, displayed in three consecutive bands of the flattened papyrus surface (I-III). Black background corresponds to regions outside the preserved writing surface or to physically absent or damaged substrate. Within the gray papyrus surface, dark strokes indicate the scholar-facing ink rendering used for papyrological inspection. The figure summarizes the transition from a sealed physical roll to a continuous inspectable writing surface. Scale bars as shown.}
  \label{fig:1}
\end{figure}

\textit{PHerc.}~1667 is the first Herculaneum papyrus to be completely unwrapped and read by virtual methods (Fig.~\ref{fig:1}). The surviving core is a remnant of earlier attempts at physical unrolling. The text appears to concern ethical theory and human moral perfectibility, with the mention of Aristocreon and other features suggesting a Stoic context (\hyperref[sec:methods]{Methods}). The sealed object is shown in Fig.~\ref{fig:1}a, while transverse and longitudinal tomographic sections reveal the dense nested architecture of the carbonized papyrus layers (Fig.~\ref{fig:1}b,c). Fig.~\ref{fig:1}d shows the corresponding 2D ink-enhanced image derived from the surface-conditioned render after virtual unwrapping: the flattened papyrus surface is displayed in three consecutive bands, with dark strokes indicating the scholar-facing ink signal used for papyrological inspection. Papyrologists localized approximately 860~cm$^{2}$ of preserved writing surface, containing the lower portion of 22~columns or column-equivalents, each comprising up to 12 lines, within the 1231~cm$^{2}$ of virtually unwrapped papyrus; black regions mark areas outside the preserved surface or physically absent or damaged substrate.

We use “complete virtual unwrapping” in a bounded geometric sense: the preserved surface is represented by a mesh whose position has been verified against the µCT volume by a human expert annotator, and is rendered in the flattened coordinate domain. We use “complete reading” in a corresponding papyrological sense: the transcription is restricted to image-supported preserved text visible in the renderings; restoration of letters corresponding to physically lost substrate is indicated by the use of square brackets in accordance with established papyrological convention; uncertain letters are marked with an underdot. The entire recovered writing surface was examined, with undecipherable passages recorded as traces and uncertainty retained in the transcription. Within this writing surface, a region of 33~cm$^{2}$ preserves ink traces that do not permit secure decipherment. The full papyrological transcription and uncertainty markings are provided in \hyperref[sec:methods]{Methods}.

\subsection*{\textit{PHerc.~Paris.}~4: volumetric validation}

\begin{figure}[H]\centering
  \includegraphics[width=0.85\linewidth]{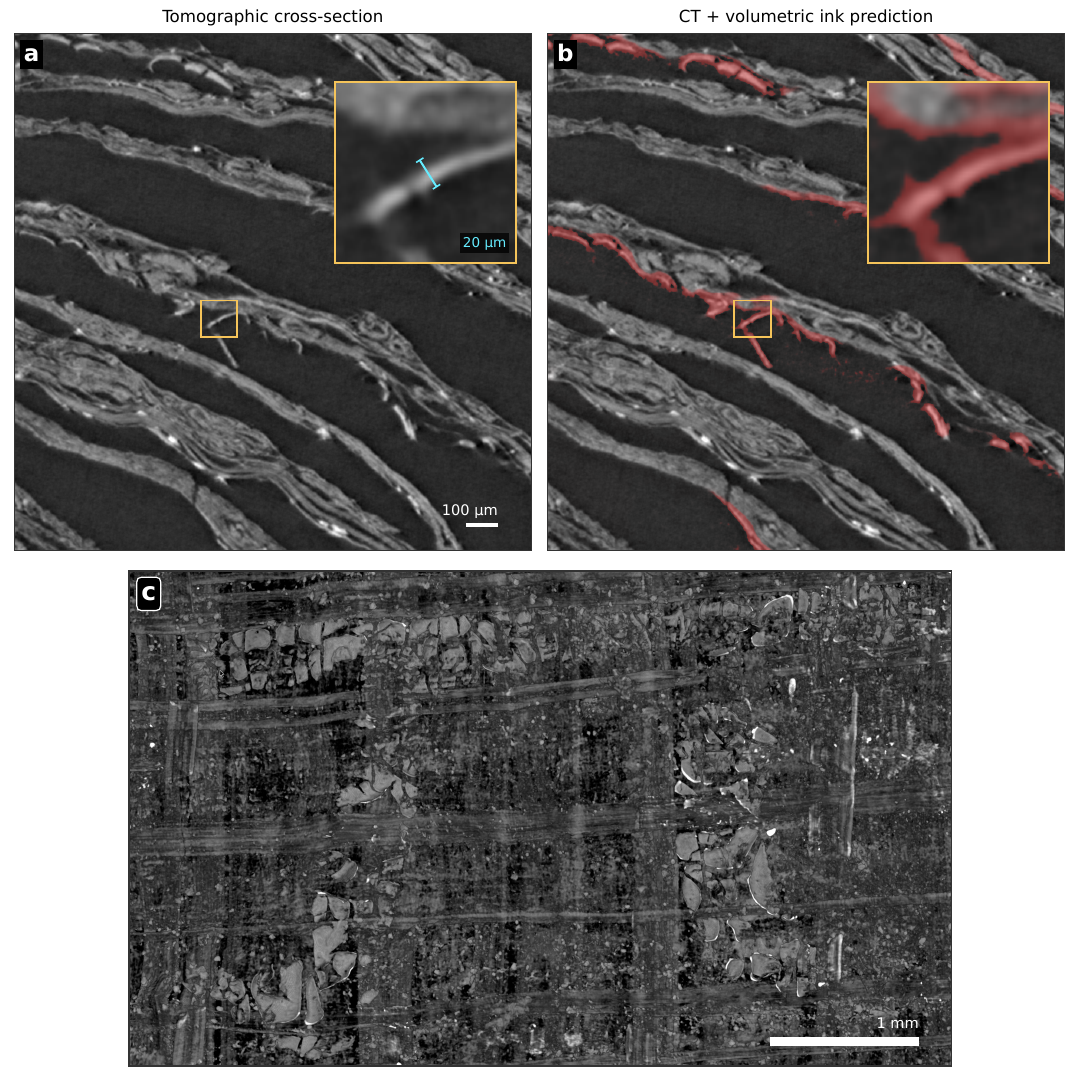}
  \caption{\textbf{Volumetric ink segmentation in \textit{PHerc.~Paris.}~4 and projection onto the unwrapped surface.} a, Native-resolution tomographic cross-section (2.4~µm isotropic voxels), with a magnified inset of a visible surface-associated deposit. The cyan ruler is a 20~µm reference marker oriented along the local in-plane normal of the deposit. b, The same CT slice overlaid in red with volumetric ink-prediction intensities; the main view and inset share the geometry and scales of a. White scale bar in a, 100~µm. c, 2D image derived from the surface-conditioned render of an ink-bearing region before projection of the volumetric segmentation, showing the underlying papyrus texture and writing traces.}
  \label{fig:2}
\end{figure}

\textit{PHerc.~Paris.}~4 tests whether surface-conditioned ink recovery corresponds to physically visible structures in the µCT volume. In the 2.4~µm high-resolution BM18 scan of \textit{PHerc.~Paris.}~4, ink-bearing strokes appear in the tomographic volume as bright, surface-associated deposits visible across several voxels (Fig.~\ref{fig:2}a,b and \hyperref[fig:ed3]{Extended Data Fig.~\ref*{fig:ed3}}).

\begin{figure}[H]\centering
  \includegraphics[width=0.85\linewidth]{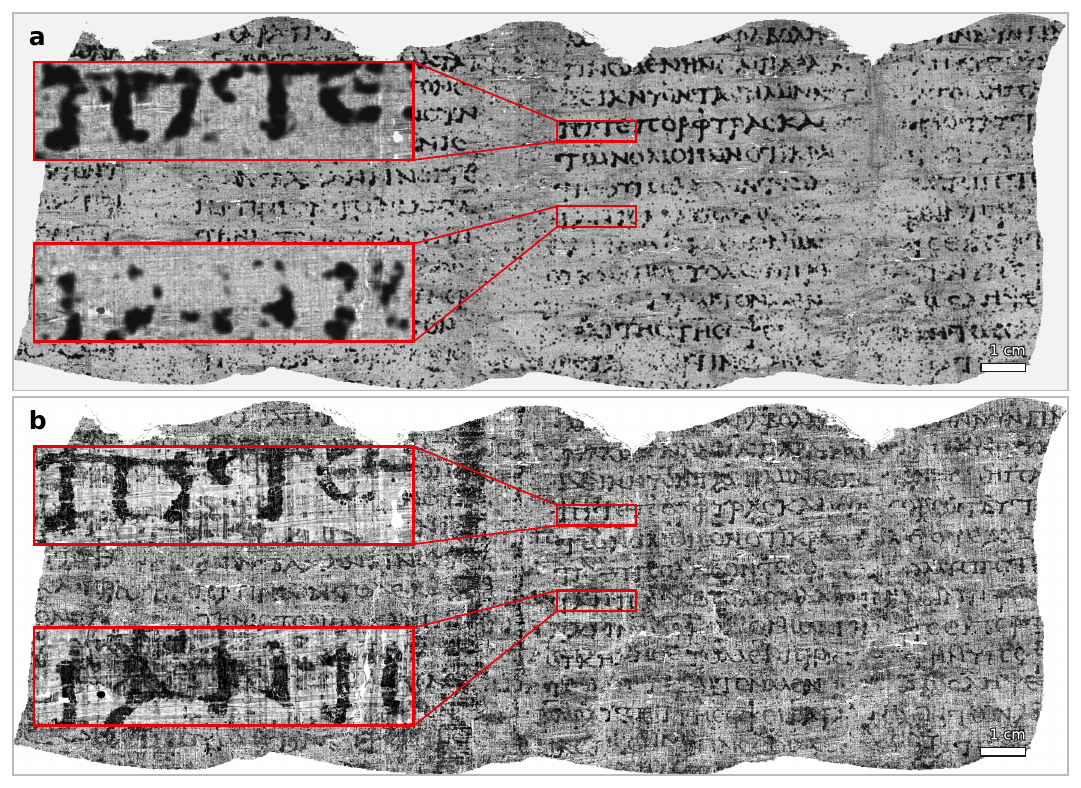}
  \caption{\textbf{\textit{PHerc.~Paris.}~4, comparison with 2023 data.} a, 2023 Grand Prize winning ink detection model\cite{ref9,ref10} on a virtually unwrapped portion of \textit{PHerc.~Paris.}~4 on the 7.91~µm pixel size volume of \textit{PHerc.~Paris.}~4 from the EduceLab-Scrolls dataset \cite{ref7}.  b, Surface rendering of the 2.4~µm data acquired at the ESRF BM18 beamline, using volumetric ink predictions to guide alpha compositing. The second confirms the reading from the 2023 Vesuvius Challenge. Zoom-in lenses highlight the difference in data quality and accurate representation of the ancient script.}
  \label{fig:3}
\end{figure}

These deposits are directly visible in the volume, can be segmented in 3D and then projected onto the flattened surface, where they coincide with the Vesuvius Challenge Grand Prize Banner (2023) region (Fig.~\ref{fig:3}), for which multi-column readings were previously published\cite{ref8,ref9,ref10}.  The correspondence between these deposits and the recovered text provides a directly inspectable physical reference for the reading. Agreement between the learned outputs complements this check (\hyperref[sec:methods]{Methods}).

\FloatBarrier

\subsection*{\textit{PHerc.}~139: identifying a sealed work}

The title of ancient bookrolls was typically written twice: at the beginning, and also at the end of the text, in the innermost wraps. These are often preserved even where the outer wraps, and with them the initial title and the opening columns were lost to the eruption, the subsequent excavation or attempts at physical opening.

\begin{figure}[H]\centering
  \includegraphics[width=0.85\linewidth]{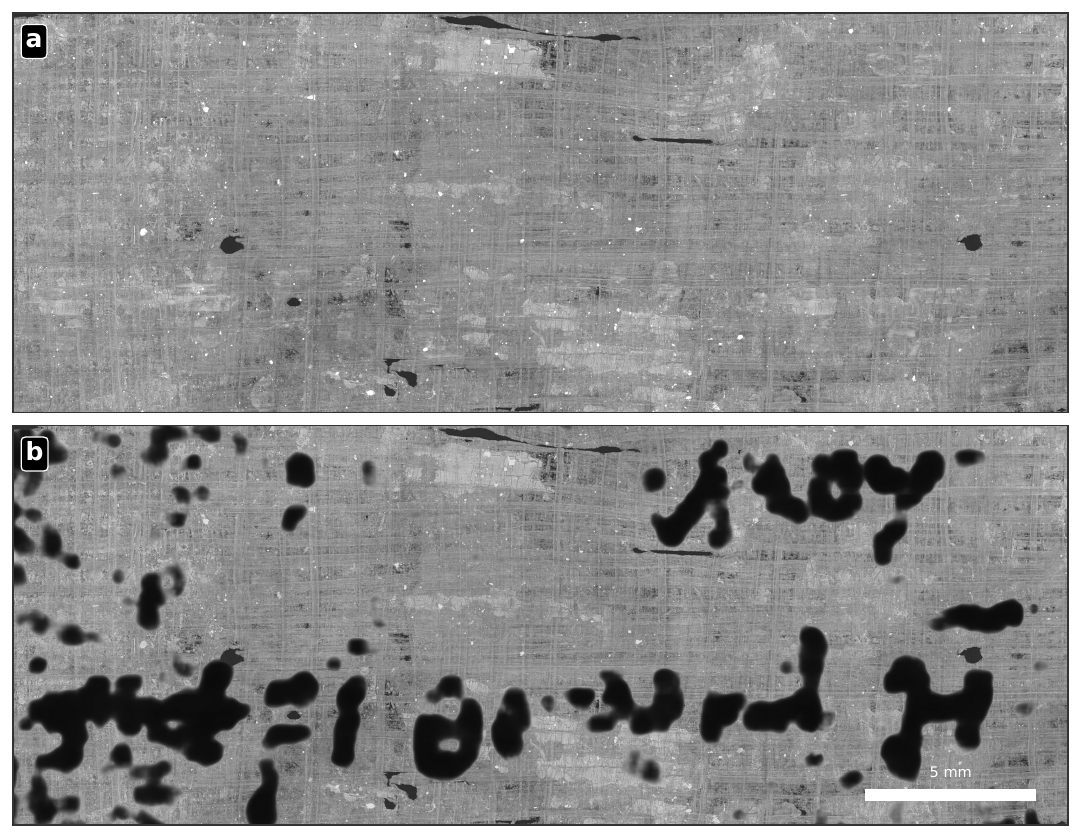}
  \caption{\textbf{Title and author-attribution evidence in \textit{PHerc.}~139.} a, 2D image derived from the surface-conditioned render of the title-bearing region of \textit{PHerc.}~139. b, The ink-enhanced signal shown in black against the papyrus texture. The broad, locally discontinuous predictions provide approximate stroke contours. The visible title and author-attribution sequence identifies the roll as Philodemus, \emph{On Gods}, Book~8. Scale bar, 5~mm.}
  \label{fig:4}
\end{figure}

\textit{PHerc.}~139 provides a separate title-identification result, although its ink predictions are less sharply defined than the examples from \textit{PHerc.}~1667 and \textit{PHerc.~Paris.}~4. The predicted strokes are broad, rounded and locally discontinuous (Fig.~\ref{fig:4}b; Extended Data Fig.~\ref{fig:sup1}), making their contours approximate guides to ink position rather than faithful outlines of the original letterforms. Papyrological interpretation of a title-bearing region of the reconstructed surface yields author, work-title and book-number evidence (Fig.~\ref{fig:4}). The title-bearing sequence is read as

\Needspace*{4\baselineskip}
\textgreek{Φ̣ι[λοδ]ή̣μου}

\textgreek{Περὶ θεῶν Η̅}

\noindent that is, “Philodemus, \emph{On Gods}, Book~8”. The overlined Η is interpreted as the book-number indication.
{}

The book number is written here on the same line as the title, despite the more commonly attested practice in titles from Herculaneum, where the title and book number appear on two separate lines. A similar layout is now documented in the final title of the unopened roll \textit{PHerc.}~172 (\textgreek{Φιλο̣δήμο̣υ̣ Πε̣ρὶ | κα̣κ̣ιῶν Α̣̅}, Philodemus, \emph{On Vices}, Book~1)\cite{ref13}. In \textit{PHerc.}~139, the title was likely written by the same scribe who copied the text.

The presence of Philodemus’ name, by far the most extensively attested Epicurean philosopher in the Herculaneum library, allows us to assign the scroll to the 1st century BC-1st century AD (with a \emph{terminus ante quem} of 79 AD). More importantly, this discovery informs us for the first time that Philodemus’ work \emph{On Gods} comprised at least eight books. Until now, only the first book had been securely identified under this title, preserved in \textit{PHerc.}~26 (\textgreek{Φιλοδήμ[ου] | Περὶ θεῶν̣ | Α̅}, Philodemus, \emph{On Gods}, Book~1), together with other theological works by the same author known under different titles\cite{reviewDangelo2025}. Building on the title identification in \textit{PHerc.}~172, this result shows how even a short recovered title can reveal the previously unknown extent of an ancient work.

In addition to the end title, the portions of text recovered thus far preserve a number of terms and expressions that provide insight into the treatise’s theological concerns:  \textgreek{χωρὶϲ προνο[ί]α̣ϲ̣} (“without providence”),  \textgreek{ἀόρατα} (“invisible entities”),  \textgreek{καὶ τὸ κατ̣ὰ̣ φύϲιν} (“and the \ldots{} according to nature”), \textgreek{καὶ πάν̣τ̣α̣ χωρὶϲ πόνων} (“and everything without troubles”), \textgreek{νοερόν} (“intellectual”). The images which supported the reading are provided in the corresponding order in Extended Data Fig.~\ref{fig:sup1}.

\AddToHookNext{shipout/after}{\flushbottom}

%% file: sections/03-discussion.tex
\section*{Discussion}

These results mark a step in the transition from a Herculaneum library known only through physically opened, fragmentary witnesses toward one that can also include virtually unwrapped texts with continuous lines and columns and identified titles and authors. Virtual unwrapping of the Herculaneum papyri is no longer limited to localized breakthroughs and partial-scroll discoveries\cite{ref7,ref8,ref9,ref10}. \textit{PHerc.}~1667 shows that scroll-scale unwrapping and reading of a still-rolled papyrus is achievable. The results from \textit{PHerc.~Paris.}~4 provide an imaging-specific payoff, because 3D segmentation of directly visible ink can ground and check surface-conditioned ink recovery. The work also opens new scholarship on previously unread texts. The title-and-attribution evidence from \textit{PHerc.}~139, following the recovery of a title from \textit{PHerc.}~172\cite{ref13}, shows that the method can identify works and authors, not only recover running prose.

This workflow does not imply that all sealed Herculaneum rolls are automatically readable. Performance depends on many factors: the general preservation state of individual scrolls; the quality of the tomographic scans; the degree of layer separation and compression; the amount of deformation suffered by the scroll; the contrast of the local ink signal; and the level of surface preservation.

Two bottlenecks remain dominant. The first is geometric. Even strong surface-prediction networks can fail in highly compressed regions, where mergers, holes and sheet switches (mistakenly leaving one layer and latching onto another, thereby breaking the continuity of any recovered writing) destabilize surface estimation. The second is radiometric. Ink evidence remains weak, varies across ink recipes and local degradation states, and is further modulated by scan conditions and diffusion of the X-rays by the samples in case of large and compacted parts. These difficulties are not independent, and their interplay helps explain the complexity of optimizing processes for readability. Better geometry produces more legible 2D images derived from surface-conditioned renders, whereas better legibility improves both expert transcription and the supervision available for ink detection.

Future gains will depend partly on improved control of effective resolution (Fig.~\ref{fig:pipeline1667}, stage~2) and on a better account of how morphology, phase effects and absorption together contribute to ink visibility (stages~2 and~4). Higher-resolution and better-tuned scans (stage~2) could reduce ambiguity in both layer separation (stage~3) and ink detectability (stage~4), especially in the most compressed regions. More robust tracing networks and hybrid neural tracers that predict or refine sheet trajectories or papyrus fibers while enforcing local surface continuity and neighboring-wrap consistency should reduce the manual burden of correction and neighbor transfer (stage~3)\cite{ref21}. The latter could be achieved by improving the accuracy of annotations used to train the machine learning models or relying on self-supervised methods. Global approaches that fit one explicit spiral model to the entire scroll\cite{scpNew1}, rather than growing the surface patch by patch, are a promising route to fully automated unwrapping (stage~3). Additional fragment-to-scroll supervision, together with conservative pseudo-label expansion on intact rolls, should strengthen ink models (stage~4) across a wider range of ink recipes without sacrificing trustworthiness.

None of these remaining challenges alters the central conclusion. With the results reported here, virtual unwrapping moves beyond isolated proof-of-concept recoveries and becomes a credible scroll-scale method for restoring texts from the still-sealed Herculaneum library. Indeed, in a single integrated workflow, we show complete reading of a virtually unwrapped Herculaneum papyrus, direct tomographic validation of ink in a still-closed roll, and title-level identification of another unopened work. The question is therefore no longer whether the hidden library can be read non-invasively, but how rapidly and comprehensively its contents can be returned to scholarship. Continued advances in imaging, surface recovery and ink detection should make it possible to approach the unopened rolls not as permanently lost objects, but as a recoverable library capable of substantially enlarging the surviving record of classical thought.

%% file: sections/05-methods.tex
\phantomsection
\section*{Methods}\label{sec:methods}

\subsection*{Provenance, custody and permissions}

The objects reported here are Herculaneum papyri presently held at the Biblioteca Nazionale di Napoli, Museo Archeologico Nazionale di Napoli, and the Bibliothèque de l’Institut de France, which conserves the \textit{PHerc.~Paris.}~scrolls on behalf of the Académie des Inscriptions et Belles Lettres. Imaging and analysis were performed under permissions granted by the relevant authorities in compliance with the applicable laws, regulations and institutional agreements governing access, transport and study of the material. The methods below follow the five stages illustrated for \textit{PHerc.}~1667 in Fig.~\ref{fig:pipeline1667}; the volumetric ink-detection branch used for \textit{PHerc.~Paris.}~4 is described within Stage~4.

\subsection*{Stage~1: Fixturing}

\Needspace*{4\baselineskip}
\textbf{Photogrammetry.}

Each sample was enclosed in a custom transport and scan case generated from a metric photogrammetric reconstruction of its exterior geometry. Photographs were acquired with a Sony \textgreek{α}6300 camera mounted on a Joby GorillaPod tripod, using an Orangemonkie Foldio360 turntable inside an Orangemonkie Foldio3 studio with additional Halo Bar illumination. ArUco fiducial markers of known side length were fixed to the turntable base. These markers are binary square fiducials whose unique identifiers and four detected corners allow robust localization and metric scaling of an image-based reconstruction\cite{ref36}.

For each stable pose, the sample was photographed over a full 360° rotation at three camera heights. The object was then repositioned under conservator supervision, including inversion or alternative footholds where the morphology permitted, and the procedure was repeated until the exterior surface was covered without blind regions. Raw photographs were converted to JPEG while preserving Exif metadata. Before photogrammetric reconstruction of the sample surface, foreground masks were generated with a fine-tuned SAM 2.1 model to isolate the scroll from the studio background, turntable and fiducial markers\cite{ref37}. The masked images were used for object reconstruction so that the surrounding setup did not contaminate the sample mesh. In parallel, the corresponding unmasked photographs were retained for the metric-scaling step: the pgs-recon pipeline\cite{ref38} detected and triangulated the ArUco markers from the full images and applied their known side length to place the reconstructed scene at physical scale\cite{ref34,ref36}.

Photogrammetric reconstruction was performed with pgs-recon, a Python pipeline built around OpenMVG for structure-from-motion and OpenMVS for dense reconstruction, mesh refinement and texturing\cite{ref34,ref35}. This produced a textured, metric-scale exterior mesh of each sample. The photogrammetry and case-generation workflow is documented in the Villa repository\footnote{\url{https://github.com/ScrollPrize/villa/commit/6847063ffdb4da898ae8d1d494ebf7d71473f509}}.

\Needspace*{4\baselineskip}
\noindent\textbf{3D case construction.} 

The mesh was then passed to the ScrollCase workflow\footnote{\url{https://github.com/ScrollPrize/villa/tree/main/foundation/scrollcase}}, which generated smoothed STL models for the sample surrogate and two complementary case shells. One shell incorporated the beamline tomograph mounting interface, and the two shells were designed to close with screws. Cases were fabricated by 3DnA s.r.l. from nylon powder using HP Multi Jet Fusion. During mounting, a conservator lined the cavity with a thin protective sheet, placed the sample into the corresponding shell, wrapped the sample with the protective material and closed the case. PTFE sheet was used for most objects. For \textit{PHerc.~Paris.}~4, which was prepared later, Japanese paper was used instead, and the case was polyurethane-coated to reduce surface roughness.

\subsection*{Stage~2: Synchrotron scanning}

\Needspace*{4\baselineskip}
\textbf{Protocol establishment.}

Tests on different samples were used to select parameters that support virtual unwrapping (Extended Data Fig.~\ref{fig:ed2}; Supplementary Table~\ref{tab:sup5}). Earlier studies of the Herculaneum papyri emphasized propagation-based phase-contrast enhancement at effective pixel sizes of 46--47~µm\cite{ref2,ref3}, with incident energy selection guided by the contrast-transfer function for the chosen propagation distance and detector sampling\cite{ref3}. The present protocol prioritizes the separation of tightly packed papyrus layers at nearly twenty-fold finer sampling (2.4~µm). Incident energy and propagation distance were therefore selected jointly to retain sufficient phase contrast while limiting scattering-induced blurring that obscures layer boundaries. Densely packed papyrus layers lead to strong beam diffusion that makes phase contrast challenging. This phenomenon, known as decoherence, blurs the projections, and impacts the visibility of small structures. In carbonized papyri, it could be linked to the presence of graphite (a very efficient decoherer) and to the natural cellular structure of the papyrus that can act as a series of very small refractive lenses and that locally deviates the beam. The solution to counter-act decoherence effect can be to reduce the propagation distance and to increase the energy. At a fixed spatial scale in the near-field regime, both changes reduce propagation-induced phase contrast, while absorption contrast remains weak for carbon-based material at hard-X-ray energies. The aim was to keep sufficient phase-contrast effect to enhance the structures of interest, without reaching the level that would result in visible decoherence effect. The production configuration selected for complete-scroll imaging was 2.4~µm isotropic voxel size, 22~cm propagation distance and 78~keV average incident energy. Finer voxel sizes improve image quality but are currently impractical for full-scroll coverage; a targeted region of interest was nevertheless acquired at a pixel size of 1.1~µm.

\Needspace*{4\baselineskip}
\noindent\textbf{Scanning.}

We used a polychromatic beam from the 1.1 T side pole of the BM18 tripole wiggler. Filtration differed between acquisitions: the \textit{PHerc.}~1667 and \textit{PHerc.}~139 scans used 2.5~mm of sapphire and 0.20~mm of silver, whereas the \textit{PHerc.~Paris.}~4 scan used 20~mm of sapphire and 0.10~mm of gold. In the gold-filtered configuration, the thin gold filter reduces transmission above its K-edge (80~keV), narrowing the effective bandwidth.

We used a scintillator-based indirect detector coupling a GAGG:Ce (Cerium-doped Gadolinium Aluminium Gallium Garnet) single crystal of 50~µm thickness with a front reflective deposit to increase its light emission efficiency. As this scintillator contains gadolinium, its X-ray stopping power increases abruptly when the energy crosses the gadolinium K-edge at 50.4~keV, making it very efficient for X-rays up to few dozens of keV above this energy. For the gold-filtered configuration, the effectively detected beam spectrum has an average of 78~keV, with enhanced detection between 50.4~keV and 80~keV. For these weakly absorbing samples, the narrowed spectrum is expected to limit beam-hardening effects. The configuration combines a filtered polychromatic beam with no focusing X-ray optics. Indeed, the only X-ray optics in the beam are sapphire and gold or silver filters, depending on the acquisition, as well as permanent polished windows (1.4~mm of diamond in the front-end and 1~mm of aluminum at the beginning of the experimental hutch).

The optic system was a tandem optic made of two photographic objectives, a Zeiss Otus 100~mm F/D 1.4 as the bottom objective, and a Nikon AF-S Nikkor 200~mm F/D 2 as the top objective. The coupling of these two objectives brings an exceptionally high numerical aperture of 0.35 that gives very high efficiency coupled with very good effective resolution. When coupled with the scintillator mentioned above, the calculated diffraction limit is of 2.72~µm for a pixel size of 2.4~µm given by the PCO edge bi-10 sCMOS camera (4.6~µm of pixel size, 85\% of quantum efficiency for the green emission of GAGG:Ce).

Data were acquired on BM18\cite{recVijayakumar2024} in helical mode\cite{recDejea2026} (Fig.~\ref{fig:pipeline1667}, stage~2B,C). To extend the transverse coverage, each sample was scanned at up to four lateral positions relative to the beam and detector, each scan sampling a partially overlapping annular region around the same rotation axis; the number of positions and of projections was chosen according to the sample diameter. For some of the scans at 1.1~µm pixel size (\hyperref[tab:ed1]{Extended Data Table~\ref*{tab:ed1}}), the field of view was instead extended with tiled mosaics. 

\Needspace*{4\baselineskip}
\noindent\textbf{Tomographic reconstruction.}

Volumes were reconstructed using the Night Rail software framework\cite{recNightRail,recDejea2026}, which integrates an HPC-oriented fork of Nabu\cite{recNabu} implementing GPU-accelerated generalized hierarchical backprojection for large helical and multi-scan BM18 acquisitions. Projections were corrected for dark current, flat field, hot pixels and detector distortion, the latter using a map calibrated independently on a slit grid and applied by bilinear resampling to both radiographs and their blending weights. Single-distance Paganin phase retrieval\cite{recPaganin2002} was then applied and the resulting blur compensated with an unsharp mask\cite{recSheppard2004}; the parameters used for each scan are listed in \hyperref[tab:ed1]{Extended Data Table~\ref*{tab:ed1}}. Intensities were then clipped to $[10^{-6},10]$ before logarithmic transformation.

In both the annular offset scans and the tiled mosaics, slow motion of the sample had to be compensated before the data could be combined. Two different strategies were used. When the sample and rotation-axis reference frame were translated together, their mutual geometry remained unchanged: adjacent acquisitions were aligned by correlating overlapping detector regions, relying on the dominant papyrus structures in the radiographs. Displacements were estimated every 5°, smoothly fitted versus projection angle and accumulated along the scan chain. The corrected radiographs were then blended before generalized hierarchical backprojection.

For tiled mosaics, the sample was instead displaced relative to the rotation axis, so each acquisition produced a separate subvolume. Slow motion appeared as a smooth, low-frequency deformation between subvolume coordinate systems and was represented through an elastically deformable common atlas. Corresponding control points were detected with Harris3D in overlapping regions; a CNN retained points belonging to the papyrus and rejected those from the support, protective case or background. The remaining matches determined the smooth displacement fields used to align and fuse the subvolumes.

The apparent rotation-axis position was also calibrated as a function of height, to account for slow variations of the effective center of rotation along the vertical direction, caused for example by a possible mechanical misalignment between the vertical translation axis and the rotation axis. 

The 32-bit reconstructions were converted to 16-bit TIFF stacks using black and white saturation levels of 0.01\%, with binned ($\times2$, $\times4$) and JPEG2000 copies archived at the ESRF\cite{dataESRF1667,dataESRF139,dataESRFParis4}. For all analyses reported here, we converted the reconstructions to unsigned 8-bit OME-Zarr\cite{ref25} with six resolution levels, with voxels outside the sample set to zero (about 20\,TB per volume); these volumes are distributed through the Vesuvius Challenge website (see \hyperref[sec:dataavailability]{Data availability}).

\subsection*{Stage~3: Virtual unwrapping}

\Needspace*{4\baselineskip}
\textbf{Recto-surface annotations and prediction.}

Recto-surface prediction was formulated as supervised 3D semantic segmentation of the papyrus writing surface $P$ in the reconstructed volume $V$. The target surface was the recto side of the sheet, defined operationally as the side facing the center of the scroll and corresponding, where visible, to the horizontal-fiber layer. Manual training labels were generated by voxelizing annotated surface meshes into binary volumetric masks. These labels were treated as approximate surface annotations rather than voxel-exact ground truth, because dense packing, damage, fraying and low local contrast make voxel-accurate manual labeling impractical at scale. In ambiguous or compressed regions, labels therefore marked the best effort continuous sheet surface suitable for subsequent tracing rather than requiring an unambiguous separation of recto and verso.

We trained a residual-encoder nnU-Net-style\cite{ref32} 3D U-Net on CT patches normalized individually by z-score. The network produced a two-channel background/surface prediction. Examples of recto-surface predictions are shown in Extended Data Fig.~\ref{fig:ed4}. The training objective used Medial Surface Recall, a custom adaptation of Skeleton Recall Loss\cite{ref33} for thin-structure segmentation, together with cross-entropy and soft Dice terms. This objective was chosen to emphasize recovery of continuous sheet traces and to reduce holes in the predicted thin surface. Training details are reported in Supplementary Table~\ref{tab:sup1}. The resulting voxelwise prediction was used only as an intermediate cue for geometric tracing: because dense predictions can contain local sheet mergers, gaps and false positives in tightly packed regions, it was converted into orientation evidence for the reconstruction of the quadrilateral mesh $M$ described below.

\Needspace*{4\baselineskip}
\noindent\textbf{Surface estimation as a quad mesh.}

The mesh $M$ that models $P$ is assembled from smaller, locally fitted patches and stored as a tifxyz quadrilateral mesh. In this custom format, the surface is represented by a regular 2D parameter grid, and each valid grid vertex stores its 3D position in $V$. The three coordinate components are saved as aligned TIFF images, x.tif, y.tif, and z.tif; adjacent valid vertices define quadrilateral mesh cells.

Automatic surface estimation began from binary volumetric predictions of the recto surface. For each of the three orthogonal slice families, predicted sheet regions were reduced to centerline traces. Each trace provided local in-plane orientation evidence for the sheet, and these directions were stored in spatial bins so that nearby sheet orientations could be queried efficiently during mesh optimization.

Each base mesh was initialized from a seed point or from an already reconstructed neighboring patch. Starting from this initial mesh, the software iteratively proposed new vertices along the boundary of the 2D mesh grid and optimized their 3D positions. The optimization jointly balanced regular mesh spacing, local smoothness, agreement with nearby orientation evidence, and proximity to the binary surface predictions, allowing the mesh to follow the predicted sheet while preserving a coherent parameterization. {}

To form larger unwrappings, a separate surface-based tracing tool combined base meshes whose seed locations and density had been chosen to ensure substantial overlap. Starting from a seed quad, this tool optimized new mesh vertices toward the consensus surface defined by the overlapping base meshes. This enabled generation of large meshes, in some cases approaching the scale of an entire scroll, with improved quality compared to the base patches. However, mergers or holes in the binary surface predictions could still lead to holes or deviations in the large mesh, particularly where the papyrus was densely packed or anomalously curved. Areas of sufficient quality were manually extracted and used as additional mesh sources.

Where a high-quality mesh already existed for one winding of a scroll, the mesh-offset tool was also employed. It traced rays from the vertices of the base mesh along the mesh normal until each ray intersected a new surface prediction. The resulting mesh was then refined using the same prediction-based mesh optimization objectives.

For meshes from all sources, manual correction was performed on the same mesh representation using an interactive annotation interface. Annotators could add control points, move vertices to corrected positions, and use push or pull operations to displace local mesh neighborhoods along the local mesh normal. The agreement of $M$ with $P$ was assessed by inspecting mesh intersections overlaid on orthogonal slices of $V$, interactively rotating the 3D view, and visualizing a CT-textured 2D grid in real time. Regions judged geometrically consistent with a single sheet were marked with an approval mask. After edits, local re-optimization propagated the correction constraints while restoring smoothness and mesh consistency. Small holes or invalid cells in the mesh grid could be repaired by local geometric inpainting where the surrounding sheet trajectory and layer identity were unambiguous. This inpainting was used only to fill gaps in the mesh and repair geometric continuity; it did not create tomographic intensity values, ink signal, or image evidence for unsupported readings.

\Needspace*{4\baselineskip}
\noindent\textbf{Flattening and rendering.}

For each recovered piece, the corrected mesh $M$ was used for final flattening and normal-direction sampling (Fig.~\ref{fig:surface1667}b--d). Valid quadrilateral cells were converted to a triangular OBJ mesh by splitting each cell into two triangles. Per-corner texture coordinates were initialized from the original surface-grid coordinates and scaled according to the surface metadata. The mesh was parameterized with SLIM\cite{ref24}, initialized from these UV coordinates. The optimization used the symmetric Dirichlet distortion energy, which is usually used in the literature as an isometric parametrization objective, with the boundary constrained to its initial UV coordinates, producing a low-distortion 2D parameter domain, with flattened coordinates $(u,v)$. The optimized UVs were shifted to non-negative coordinates and written back to OBJ.

The optimized mesh was rasterized back into tifxyz form in the flattened UV domain. For each output pixel, the corresponding 3D surface position was computed by barycentric interpolation within the containing triangle. This inverse map through $M$ assigns each flattened coordinate $(u,v)$ its 3D reference position in $V$.

For rendering, the software evaluated the flattened surface on a regular 2D canvas. At each canvas pixel, the renderer computed the corresponding 3D surface point and local surface normal. A stack of 65~samples was taken along this normal direction, centered on the reference surface. For sample index $k$, the normal offset was
\begin{equation*}
d_k = (k - (N - 1)/2)\Delta,\quad k=0,\ldots,N-1,
\end{equation*}
with $N=65$ and $\Delta=1$ voxel. This gives offsets from -32 to +32~voxels; at 2.4~µm isotropic voxel sampling, the distance between the first and last samples is $64 \times 2.4 = 153.6$ µm. Tomographic intensities were sampled at each off-surface position by trilinear interpolation from the eight neighboring voxels. Samples outside the valid volume bounds were assigned zero. The resulting warped 3D CT chunk is the surface-conditioned render $S(u,v,d_k)$ (Fig.~\ref{fig:surface1667}d). Its two in-plane coordinates follow the flattened mesh, and its depth coordinate samples the neighborhood along the local normal in $V$. Each fixed-depth slice is a 2D render; the central slice ($d_k=0$) samples $V$ on $M$. A selected depth range of $S$ forms the input to the ink-detection models described below.

\subsection*{Stage~4: Ink detection}

\Needspace*{4\baselineskip}
\textbf{Fragment registration for ink labels.}

The first supervised ink labels came from detached scroll fragments, whose writing was directly exposed by historical attempts at mechanical opening. Infrared photographs, acquired with a MegaVision multispectral system equipped with a 50-megapixel E7 camera back, a UV–IR apochromatic lens and four UV–IR LED light panels, were aligned to CT-based surface representations\cite{ref7,ref16,seales2023cerc}.

These fragments provide the direct-visibility labels mentioned in the Technical Strategy and serve as reference for ink location in the surface plane. Because the writing is directly exposed, infrared photography can capture it with high contrast and visibility, isolating the ink signal that is otherwise difficult to detect. A corresponding CT-based surface is generated through an orthographic projection, casting rays onto the digital volume in the same manner that the camera images the fragment in the real world. The infrared image is then warped to align with the visible features of the digital volume. Extended Data Fig.~\ref{fig:ed5} shows a registered infrared image and its CT-based surface representation.

After registration, the resulting labels were mapped into the same flattened 2D coordinate domain $(u,v)$ associated with $S$ used for training and inference. This permits the detector to learn from physically exposed writing while being deployed on surfaces extracted from sealed scroll interiors.

\Needspace*{4\baselineskip}
\noindent\textbf{Ink detection: from fragments to sealed scrolls.}

For \textit{PHerc.}~139 and \textit{PHerc.}~1667, each detector used local 3D chunks of $S$ as input. During training, these chunks were aligned with 2D label regions in $(u,v)$.
Initial training used paired virtually unwrapped CT surface volumes and registered infrared ink labels from \textit{PHerc.}~9B, \textit{PHerc.}~343P and \textit{PHerc.}~500P2. The infrared images accurately localize the ink signal in the surface plane, but because they capture only a top-down view of each fragment, they provide no information about the ink's position in depth. The ink detectors use 3D ResNet\cite{ref42} encoders with U-Net-style\cite{ronneberger2015unet} decoding to predict ink in the 2D coordinate domain $(u,v)$, in which the infrared labels are defined. The decoder may operate in 2D after depth pooling or in 3D before depth reduction; the configuration used for the \textit{PHerc.}~1667 label-coverage ablation and the model used for the \textit{PHerc.}~139 images in Extended Data Fig.~\ref{fig:sup1} are described in Supplementary Tables~\ref{tab:sup2} and~\ref{tab:139canonical}, respectively.

The ink detector sees only local 3D chunks of $S$, constraining the task to local CT texture and morphology and limiting the scope for predictions driven by linguistic or word-level priors. We chose an input size of 256~$\times$~256 pixels in the surface plane, which roughly corresponds to 614~$\times$~614~µm at 2.4~µm sampling (Extended Data Fig.~\ref{fig:ed6}). This input window was chosen to emphasize local CT structure and restrict the context available about the surrounding text.

The fragment-trained model detected initial ink traces on \textit{PHerc.}~139 and \textit{PHerc.}~1667 before any training on those scrolls, providing a pretrained backbone for scroll-specific adaptation.

Adaptation used iterative pseudo-labeling based on the model’s own predictions\cite{zou2018selftraining} refined by an expert (Extended Data Fig.~\ref{fig:ed7}). Starting from an initial set of labeled tiles, we fine-tuned the model and ran inference across the surface-conditioned renders $S$ of the scroll. Regions in which the model revealed additional candidate ink were cropped into new training tiles, with pseudo-labels derived from its predictions. Repeating this process expanded the dataset and improved the consistency and visibility of the ink signal for papyrological review. For \textit{PHerc.}~1667, this procedure was applied at larger scale using three columns for iterative labeling and five iterations. We trained with a mixture of Dice loss and binary cross-entropy loss with label smoothing to limit overfitting to the signal and to reduce the impact of label noise\cite{zhang2022segmentation}.

Additionally, the same procedure was applied to a scanned region of interest in \textit{PHerc.}~1667 with a pixel size of 1.1~µm, enabling the reading of previously illegible portions of the text and improving their legibility.

A label-coverage ablation (Extended Data Fig.~\ref{fig:ed7}) compared matched models whose ResNet3D-50 encoders were initialized from the same weights pretrained on the combined Kinetics-700 and Moments in Time video datasets\cite{ref42,kataoka2020mega} and trained for the same number of steps, using label sets from successive pseudo-labeling rounds. For the ablation, successive label sets were developed on segment $\ell_2$. The matched models were assessed qualitatively on segment $\ell_5$, which was held out throughout training and has no reference ink labels. Across the successive label sets, additional text became visible in the held-out column.

\Needspace*{4\baselineskip}
\noindent\textbf{Volumetric ink segmentation in \textit{PHerc.~Paris.}~4.}

Because the 2.4~µm-voxel volume $V$ of \textit{PHerc.~Paris.}~4 preserves enough ink–substrate contrast for direct visual identification of ink, we trained a 3D residual U-Net to predict per-voxel ink probability in $V$ from 256$^{3}$-voxel patches. A custom 3D DINOv2\cite{ref22} representation was used to derive coarse 3D ink labels. Architectures and training procedures are described in Supplementary Tables~\ref{tab:sup3} and~\ref{tab:sup4}.

Dense supervision was obtained primarily from the \textit{PHerc.~Paris.}~4-finetuned 3D DINOv2 embedding space. An expert selected a small set of visibly inked voxels in axis-aligned slices, and their DINO patch embeddings were averaged into a normalized ink prototype. Cosine similarity to this prototype produced a dense 3D ink-likeness map, which provided the main label signal.

To make these DINO-derived labels conservative, we combined two complementary signals: the DINO ink-likeness map and the output of a separately trained volumetric ink detector. This detector was trained from 2D annotations made on images derived from $S$, mapped into $V$ through $M$ and extruded along the local normal to $M$ to give them an approximate thickness, and was therefore useful mainly as a coarse support constraint: it reduced spurious DINO responses, while the DINO map supplied the finer localization cue. In practice, the two maps were multiplied and thresholded, with low-intensity air and void voxels masked out from the second guided round onward so that labels could only occur near papyrus material.

A student detector was trained on these combined labels, then the procedure was repeated using the guided student as the coarse support model. This was intended to refine the DINO-derived supervision while preserving the DINO ink prototype as the primary localization cue. Finally, we self-distilled the DINO-guided detector by generating dense labels from two guided-detector checkpoints under test-time augmentation, retaining material and intensity gates. This final step was intended to sharpen boundaries and suppress residual artifacts. The predicted ink volume was subsequently projected onto flattened surfaces.

%% file: sections/05b-methods-validation.tex
\subsection*{Stage~5: Papyrological interpretation}

\Needspace*{4\baselineskip}
\textbf{Validation and papyrological review.}

For \textit{PHerc.}~1667, the flattened and ink-enhanced pieces were then merged into one coherent scroll view for papyrological interpretation (Fig.~\ref{fig:pipeline1667}, stages~4--5). The \textit{PHerc.}~1667 transcription was produced and interpreted at column level by nine qualified papyrologists. Reviewers inspected the flattened renderings, ink-enhanced outputs and geometric context of the sampled surface. Accepted readings met three criteria: (i) the sampled surface was geometrically tight to the papyrus layer in the CT volume; (ii) letterforms remained consistent across repeated renderings or inference passes; and (iii) the transcription was independently endorsed or resolved by consensus.

Uncertainty in the readings was retained in the transcription.
Papyrological review could prompt reinspection and refinement of surface geometry, rendering or ink enhancement (Fig.~\ref{fig:pipeline1667}, dashed feedback paths from stage~5 to stages~3 and~4).

%% file: sections/06-transcription.tex
\subsection*{Complete \textit{PHerc.}~1667 transcription and translation}

\textit{PHerc.}~1667 is today a portion of a scroll that is incomplete in both its height (8~cm out of the usual 19--24~cm of a full roll) and diameter (2~cm out of 4--6~cm for a full roll). Repeated attempts to physically open \textit{PHerc.}~1667 were unsuccessful. Archival records document an unsuccessful attempt in the nineteenth century, after which the papyrus was classified as a \emph{midollo}, the compact inner core of a carbonized roll from which the outer layers had been removed. Another attempt to open the papyrus was initiated in November 1969 by the conservator A. Fackelmann, but it was unsuccessful and was therefore discontinued\cite{ref40,fackelmann1970}. In the 1980s, the papyrus was subjected to the Oslo method, a mechanical opening technique developed for Herculaneum scrolls, which yielded ten fragments from the outer layers\cite{ref41}. Although a small number of isolated letters could be identified, overlapping sheets prevented the reading of continuous text, and the roll was consequently classified as unreadable. These interventions substantially altered the physical state of the artifact, reducing the diameter of the surviving roll from 4.9 to 2~cm, and its weight from 14~g to approximately 6~g.

Following virtual unwrapping, \textit{PHerc.}~1667 has been discovered to preserve the lower portions of the final columns of a philosophical treatise. The title, which in ancient bookrolls was typically written also at the end of the text, has not been recovered and may have been located in the now-lost upper portion of the roll. Consequently, neither the authorship nor the title of the work can be established with certainty. Despite the fragmentary state of the manuscript, the good legibility of several consecutive lines enables the identification of the principal themes addressed in the surviving portion of the text. The surviving text appears to focus primarily on ethical theory, likely in connection with the moral perfectibility of human beings, a central concern of ancient philosophy and a defining objective of Hellenistic philosophical traditions in particular. Moreover, a number of lexical and conceptual features, together with the mention of the Stoic philosopher Aristocreon, the nephew and disciple of Chrysippus, in what appears to be the final preserved column, point toward a Stoic context.

The paleographical and textual evidence—the mention of Aristocreon, whose period of activity is generally placed around the end of the 3rd century BC—supports dating \textit{PHerc.}~1667 to the late 3rd century BC or beginning of 2nd century BC.

A comprehensive interpretation and commentary lie beyond the scope of the present paper and will be the subject of a separate papyrological study. The text presented here should be understood as a preliminary transcription of the Greek text, accompanied by an English translation, intended to support the validation of the virtual unwrapping and ink detection processes.

\par
\raggedbottom
\addvspace{\baselineskip}
\noindent\textbf{Coll. 1-4}\par\nobreak\vspace{\baselineskip}
\begin{transpair}
\begin{transcription}
            \emph{traces}   
\end{transcription}
\end{transpair}
            
\noindent\textbf{Col.~5}\par\nobreak\vspace{\baselineskip}
\begin{transpair}
\begin{transcription}
            \emph{lost margin}
            ll. 1-2 \emph{traces}
            ]  ̣  ̣  ̣  ̣ τομ  ̣[
            ]τ̣ί̣θ̣εϲθαι ψ̣ευ̣  ̣[
 5         ]ϲ̣ τὸ ὅμοιον  ̣  ̣ [
\end{transcription}

{\footnotesize 3-4 \textgreek{τὸ μὴ̣ [ϲυγκα|τα]τ̣ί̣θ̣εϲθαι ψ̣ευ̣δ̣[ῶϲ} or \textgreek{τὸ μὴ̣ [ϲυγ|κατα]τ̣ί̣θ̣εϲθαι ψ̣ευ̣δ̣[ῶϲ}?}
\transversion
« … the similar …»
\end{transpair}
\par

\noindent\textbf{Col.~6}\par\nobreak\vspace{\baselineskip}
\begin{transpair}
\begin{transcription}
            \emph{lost margin}
            l. 1 \emph{traces}
            ll. 2-4 \emph{lost}
            l. 5 \emph{traces}
            ] ὁ̣ρμὰϲ περι[
            ]  ̣ϲ τομηϲυ  ̣[
            ]θεϲ̣θ̣αι τῶι   ̣[  
\end{transcription}

{\footnotesize 7-8 \textgreek{τὸ μὴ ϲυγ̣[κα|τατί]θεϲθαι} or \textgreek{τὸ μὴ ϲυγ̣[κατα|τί]θεϲθαι}?}
\transversion
« … impulses … »
\end{transpair}
\par

\noindent\textbf{Col.~7}\par\nobreak\vspace{\baselineskip}
\begin{transpair}
\begin{transcription}
            \emph{lost margin}
            ll. 1-6 \emph{traces} 
              ̣  ̣  ̣  ̣  ̣  ̣  ̣ψε  ̣[  ̣  ̣  ̣  ̣  ̣  ̣]
            ϲ̣ει̣  ̣ προϲ̣γ̣  ̣[  ̣  ̣  ̣  ̣  ̣  ̣]
              ̣  ̣  ̣  ̣  ̣το φυϲ̣ι̣[  ̣  ̣  ̣  ̣  ̣  ̣]
10        τ̣ηι̣ ἀπροπτ̣[  ̣  ̣  ̣]  ̣  ̣  
\end{transcription}

{\footnotesize 10 \textgreek{ἀπροπτ̣[ωϲί]α̣ι̣}?}
\transversion
« … deliberate(ness/ly) … »
\end{transpair}
\par

\noindent\textbf{Col.~8}\par\nobreak\vspace{\baselineskip}
\begin{transpair}
\begin{transcription}
            \emph{lost margin}
            ll. 1-5 \emph{traces}
              ̣α̣δε  ̣  ̣  ̣[  ̣  ̣  ̣  ̣  ̣  ̣  ̣]
            l. 7 \emph{traces}
            κ  ̣  ̣  ̣  ̣η  ̣[  ̣  ̣  ̣  ̣  ̣  ̣  ̣]
            τιν  ̣  ̣ε  ̣[  ̣  ̣  ̣  ̣  ̣  ̣]με-
10        ν̣ο̣ϲ παρα  ̣  ̣  ̣[  ̣  ̣  ̣]  ̣
            ἀλλα  ̣  ̣  ̣  ̣υ  ̣  ̣  ̣  ̣  ̣
\end{transcription}
\transversion
\end{transpair}
\par

\noindent\textbf{Col.~9}\par\nobreak\vspace{\baselineskip}
\begin{transpair}
\begin{transcription}
            \emph{lost margin}
            [  ̣  ̣  ̣  ̣  ̣  ̣  ̣  ̣  ̣]  ̣  ̣  ̣[1-2]
            [  ̣  ̣  ̣  ̣  ̣  ̣  ̣  ̣]  ̣  ̣ει̣εν̣[1-2]
              ̣  ̣  ̣  ̣  ̣[  ̣  ̣] κ̣α̣θ’ ὅϲον
            δ̣εξα  ̣[  ̣  ̣  ̣  ̣]  ̣τ  ̣ϲ ἐ̣ϲ̣-
 5         τιν   ̣  ̣  ̣  ̣[  ̣  ̣  ̣  ̣  ̣]  ̣  ̣
            ὑποκατα  ̣[  ̣  ̣  ̣  ̣  ̣  ̣  ̣]
            τοϲ τον[  ̣]  ̣  ̣  ̣  ̣  ̣υ
            το̣ν ἄνθ̣ρ[ωπον   ̣  ̣  ̣]ειν
            αὐτὴν ἢ ϲχεῖν̣   ̣[  ̣]  ̣  ̣[  ̣  ̣]
10        ναι ὅτι δὲ κ̣α̣  ̣  ̣  ̣  ̣οιϲ
\end{transcription}

{\footnotesize 9 \textgreek{λαβ]εῖν} Scalas}
\transversion
« … so far as … human being ... this or to have/get … that/what … »
\end{transpair}
\par

\noindent\textbf{Col.~10}\par\nobreak\vspace{\baselineskip}
\begin{transpair}
\begin{transcription}
            \emph{lost margin}
            [  ̣  ̣  ̣  ̣  ̣  ̣  ̣  ̣]  ̣  ̣  ̣[  ̣  ̣  ̣]
            τ̣ε̣κ̣  ̣  ̣[  ̣  ̣]  ̣  ̣  ̣  ̣  ̣  ̣  ̣
              ̣  ̣  ̣  ̣  ̣  ̣[  ̣  ̣  ̣]  ̣  ̣  ̣  ̣  ̣  ̣
              ̣  ̣δετ  ̣[  ̣  ̣  ̣  ̣  ̣  ̣  ̣]  ̣  ̣ 
5          ϲ̣υ̣νκριϲ  ̣[  ̣  ̣  ̣  ̣  ̣  ̣  ̣]  ̣ 
            νη̣ϲ πο  ̣  ̣[  ̣  ̣  ̣  ̣  ̣  ̣ϲ]τά-
            ναι προϲ̣ῆ̣κον ἐπὶ τὸ
            πᾶν̣ ἔ̣τι   ̣ο  ̣[  ̣  ̣  ̣]η̣ρου ἔϲ-
            ται δ̣έοϲ καὶ εν  ̣  ̣  ̣ωι
10        τὸ μέγα καὶ̣ μ̣ακρὸν
\end{transcription}

{\footnotesize 3-4 perhaps traces of a \emph{diplē} 8 \textgreek{]η̣ρου} or \textgreek{]π̣ρου}, \textgreek{τ̣οῦ̣ [πον]η̣ροῦ}? 9 \textgreek{δ̣έοϲ} or \textgreek{θ̣εόϲ}? \textgreek {ἐν α̣ὐ̣τ̣ῶι} (Hammerstaedt) or \textgreek {ἐν π̣ό̣ν̣ωι}?}
\transversion
« … that befits/belongs on the whole still … fear will be [of? …] and … the great and long … »
\end{transpair}
\par

\noindent\textbf{Col.~11}\par\nobreak\vspace{\baselineskip}
\begin{transpair}
\begin{transcription}
            \emph{lost margin}
            [  ̣  ̣  ̣  ̣  ̣]κ̣α̣ι̣  ̣[  ̣  ̣  ̣  ̣]  ̣  ̣
            [  ̣  ̣  ̣  ̣  ̣]  ̣  ̣ι̣θ̣  ̣  ̣ κ̣α̣τ̣' α̣ὐ̣-
            τ̣ὸ̣ν̣ [  ̣  ̣  ̣]α̣ϲτάτων   ̣  ̣  ̣
              ̣  ̣  ̣  ̣[  ̣  ̣  ̣]  ̣τηϲιν καὶ τὴν
 5         ὁρμ[ὴν   ̣  ̣  ̣  ̣  ̣  ̣] κ̣ατὰ
            τουτ̣[  ̣]  ̣  ̣  ̣  ̣  ̣  ̣  ̣  ̣ι̣ϲηϲ
            πρὸϲ [δ]ὲ τ̣ὰ̣ οὕτ̣ω̣ϲ ἕκαϲ-
            τα πο  ̣  ̣  ̣ω̣ϲ̣ πεφύκα-
            μεν καὶ πρὸϲ τὸ ταῦ-
10        τα ποιεῖν ἃ φαίνειν
\end{transcription}

{\footnotesize 3 \textgreek{τ̣ο̣ν̣} or \textgreek{τ̣ω̣ν̣} 4 \textgreek{ἐ̣[πιζ]ή̣τηϲιν}? 8 \textgreek{πον̣η̣ρ̣ῶ̣ϲ̣}? 9-10 \textgreek{ταῦ|τα} or \textgreek{ταὐ|τά}}
\transversion
« … according to him/it ... and the impulse … For/toward each of these things in this way … we are by nature … and for/toward doing these things that … to seem … »
\end{transpair}
\par

\noindent\textbf{Col.~12}\par\nobreak\vspace{\baselineskip}
\begin{transpair}
\begin{transcription}
            \emph{lost margin}
            [  ̣  ̣  ̣  ̣]ε  ̣  ̣  ̣ο[  ̣  ̣]  ̣κα  ̣  ̣  ̣
            [  ̣  ̣]  ̣ου  ̣  ̣ο̣ϲ̣ ἐμ̣φαίνει.
            [κοι]ν̣ότατα δ’ ἀνθρώ̣-
            [πο]ι̣ϲ̣ καὶ θηρϲὶν τάδε
 5         [  ̣  ̣  ̣  ̣  ̣  ̣ τ]έ̣κνα   ̣  ̣  ̣  ̣  ̣
              ̣[  ̣  ̣  ̣  ̣  ̣  ̣]κ̣αι ταϲ  ̣α̣  ̣
            [1-2] πολὺ δὲ μάλιϲτα τὰ̣  
            κοινότατα ἕκαϲτα
            ταῦτα ϲυνίϲτηϲιν
10        `  ̣´ηφ  ̣  ̣  ̣  ̣  ̣αι γὰρ ἀναγ-
\end{transcription}

{\footnotesize 2 \textgreek{ἐμ̣φαίνει} or {π̣ρ̣ο̣ϲ̣εμ̣φαίνει} 5 \textgreek{τ]έ̣κνα} Rashed 10 perhaps \textgreek{`μ̣´ὴ φ}}
\transversion
« … shows. Very common ... to men and beasts these … offspring ... And above all, each of the most common things constitutes these [\emph{or} these most common things each constitute] … For, [necessity? necessary?] … »
\end{transpair}
\par

\noindent\textbf{Col.~13}\par\nobreak\vspace{\baselineskip}
\begin{transpair}
\begin{transcription}
            \emph{lost margin}
            φυϲικὴν ὑπ[  ̣  ̣  ̣  ̣]  ̣  ̣  ̣
            διὸ καὶ οὐ  ̣  ̣  ̣  ̣  ̣  ̣  ̣  ̣  ̣
            κατὰ τὴν   ̣π  ̣β  ̣  ̣  ̣  ̣
            ταύτην τα  ̣  ̣κ[  ̣]  ̣  ̣  ̣
 5         [  ̣  ̣  ̣]  ̣α εὑρεθήϲετ[αι]
            καὶ οἱ βίοι προάξ̣ου̣[ϲιν]
            οὐθέν, ἡμῶν οὔτ̣ε τῆϲ
            ἡδονῆϲ οὔτε τοῦ πό-
            νου προϲδεομένω̣ν.
10       ὡϲαύτωϲ δὲ καὶ τ  ̣
\end{transcription}

{\footnotesize 1 perhaps \textgreek{ὑπ[όϲτα]ϲ̣ι̣ν̣} Hammerstaedt 3 perhaps \textgreek{ἐ̣πι̣βο̣λ̣ή̣ν̣}}
\transversion
« … natural … therefore also … according to the … this … will be found, and lives will make no progress whatsoever, as we have no need for either pleasure or pain. In the same way, also ... »
\end{transpair}
\par

\noindent\textbf{Col.~14}\par\nobreak\vspace{\baselineskip}
\begin{transpair}
\begin{transcription}
            \emph{lost margin}
            [  ̣  ̣  ̣  ̣  ̣  ̣  ̣  ̣]  ̣  ̣α̣τα
            τ̣ι αὐτο[τ]ελ̣ὲϲ καὶ ἐλ-
            λεῖπον, οὕτωϲ̣ βούλο̣-
            μα̣ι̣ λ̣έγειν α̣  ̣  ̣  ̣  ̣  ̣  ̣  ̣  ̣
 5         π  ̣  ̣  ̣  ̣  ̣  ̣π̣  ̣  ̣ τέλεον
            ἐλλείπ̣ειν τ̣[ὰ] ἀριϲτ̣ε̣ρ̣ά̣,
            καὶ ὡϲαύτ[ωϲ] ἔ̣χ̣ειν
            ἐπὶ τῶν δεξιῶν με-
            ρῶν πρὸϲ τὰ ἀριϲτερά.
10        γίνεται δ’ ὁ πλεοναϲ-
            μὸϲ κατὰ τὴν ὁρμὴν
\end{transcription}

{\footnotesize 1-2 perhaps \textgreek{κα̣τά | τι} 7 \textgreek{ἔ̣χ̣ειν} Rashed}
\transversion
« … something complete in itself and [\emph{or} also] lacking, in this way I want to say that … complete … the left-hand (parts) are lacking, and things stand in the same way for the right parts in relation to the left ones. There is an excess in the impulse … »
\end{transpair}
\par

\noindent\textbf{Col.~15}\par\nobreak\vspace{\baselineskip}
\begin{transpair}
\begin{transcription}
            \emph{lost margin}
            [  ̣  ̣  ̣]  ̣  ̣ο̣υ̣κ̣α̣  ̣  ̣[  ̣  ̣  ̣]
            γ̣  ̣  ̣υο̣ν̣τ̣οϲ καὶ τῶν
            π̣αραπ̣λ̣η̣ϲίων πάν-
            των. καθ’ οἷον γὰρ εἶδοϲ
 5         α̣ἱ ὁρμαὶ πεφύκαϲιν
            γίνεϲθαι, κ̣α̣τὰ τοῦ-
            τό ἐϲτι̣ν τὸ ἀνελλι-
            πέϲ, ὥϲτε μηθὲν ἐπι-
            ζητε`ῖ´ν ἔτι, ἀλλὰ ἀ-
10        ναπ̣ε̣π̣ληρῶϲθαι κα-
            τὰ πᾶν ὡϲ ἂν κατα
\end{transcription}

{\footnotesize 1-2 \textgreek{κα]|τ̣α̣λ̣ύο̣ν̣τ̣οϲ}? 11 perhaps \textgreek{κατὰ || [τὴν ἐπιϲτήμην} Hammerstaedt}
\transversion
« … and of all similar things. For there will be what lacks nothing with respect to this kind, according to which impulses exist by nature, so as not to seek anything any longer, but to be completed in every respect as if … »
\end{transpair}
\par

\noindent\textbf{Col.~16}\par\nobreak\vspace{\baselineskip}
\begin{transpair}
\begin{transcription}
            \emph{lost margin}
                                        γέ-]
            γονε̣ν, δ̣ι̣[ὰ τὴν φρό]νη-
            ϲ̣ι̣ν̣ ἐ̣γ̣γίζουϲ̣ι̣ τῆϲ
            ϲυντελείαϲ. ἀ[π]ὸ̣ τού-
5          των δὲ μεταβα̣ί-
            ν̣ουϲα ἐπὶ τὸ μι̣κ̣ρ̣όν,
            τὰ μὲν καθ’ αὑτὴν
            ἐν ἡμῖν ϲυντετέλε-
            κεν πάντα οὐ δ̣υ̣να-
10        μένη ϲ̣υμπληρῶ\{ι\}-
            ϲαι τὴν φύϲ̣ιν, πα-
            ρέδωκε δὲ ὑπὸ πρα-
\end{transcription}

{\footnotesize 12 \textgreek{ὑπὸ πρα||[γμάτων} or \textgreek{ὑπὸ πρά||[ξεωϲ}}
\transversion
« … happened, by means of wisdom they approach completion. Moving from these things to the small [detail?], it [a τέχνη?] has accomplished within us all that pertains to it, even though it cannot fully complete [our?] nature, and it provided … »
\end{transpair}
\par

\noindent\textbf{Col.~17}\par\nobreak\vspace{\baselineskip}
\begin{transpair}
\begin{transcription}
            \emph{lost margin}
            [  ̣  ̣  ̣]  ̣οι ζητήϲ̣ο̣-
            μέν τι, ἀλλ’ οὐχ ἕξο-
            μεν ἐὰν τρόπον τι-
            νὰ ἑαυτῶν ἀπ̣οϲτῶ̣-
 5         μεν καὶ τῆϲ φύϲεωϲ
            ἡμῶν, ἔτι τε ὃν τρό-
            πον ἂν ῥηθείηϲαν αἱ
            λοιπαὶ τέχναι κατά
            τι μὲν τέλειαι εἶναι,
10        κατά τι δὲ ἐνλείπε`ι´ν,
            τῆϲ φρονήϲεωϲ δι-
\end{transcription}
\transversion
« … we will inquire into something, but we will not grasp it, if in some way we depart from ourselves and from our own nature, and further, [this would be] in the same way as the remaining arts may be said to be perfect in one respect, but to be deficient in another respect, wisdom … »
\end{transpair}
\par

\noindent\textbf{Col.~18}\par\nobreak\vspace{\baselineskip}
\begin{transpair}
\begin{transcription}
            \emph{lost margin}
                                περὶ ἐ-]
            κείνην τ̣ὴ̣[ν φ]ρόνη-
            ϲιν οὖϲαν, καὶ περὶ αὐ-
            τὴν εἶναι, τοῦτον δὲ
 5         περὶ τὰϲ βαναύϲουϲ
            ὄντα τέχν̣α̣ϲ̣ ἀ̣πολε-
            λεῖφθαι πολὺ τοῦ τοι-
            ούτου καὶ ὥϲπερ χω-
            λὸν τὸ τεχνικὸν ἔ-
10        χειν καὶ ἐνλελειμ-
            μένον, τοιοῦτό τι μοι
            δοκεῖ καὶ ἐπὶ τῶν
\end{transcription}
\transversion
« [If we say that] ... being wisdom about that [\emph{or} being ... about that wisdom], is also about it, whereas this [person?] that deals with banausic arts is far behind from such a thing and has technical competence that is, so to speak, lame and lacking, something similar seems to me also in regard to … »
\end{transpair}
\par

\noindent\textbf{Col.~19}\par\nobreak\vspace{\baselineskip}
\begin{transpair}
\begin{transcription}
            \emph{lost margin}
            [  ̣  ̣  ̣  ̣]  ̣  ̣[  ̣  ̣  ̣  ̣  ̣  ̣  ̣  ̣]
              ̣[  ̣]  ̣δ̣εῖϲθαι μη[δ]ε-
            μιᾶϲ. ἐκπονηθέν̣-
            των μέν̣τοι̣ ἡμῶν
 5         διὰ τῆϲ ζητήϲεωϲ
            ἢ τῆϲ μαθήϲεω̣ϲ̣, κα-
            τ’ οὐθὲν ἔτ̣ι λειφθηϲό-
            μεθα αὐτῶν ὁμοίωϲ
            τέλεα τὰ καθήκον-
10        τ’ αὐτοῖϲ ποιοῦντεϲ
            καὶ τὴν αὐτὴν ἔχον-
            τεϲ αὐ̣τοῖϲ φρόνηϲιν
\end{transcription}
\transversion
« … need none. Having certainly been instructed/completed through research or learning, we will no longer be inferior to them in any respect, accomplishing the appropriate actions as they do and possessing the same wisdom as they do … »
\end{transpair}
\par

\noindent\textbf{Col.~20}\par\nobreak\vspace{\baselineskip}
\begin{transpair}
\begin{transcription}
            \emph{lost margin}
              ̣[  ̣  ̣  ̣  ̣]  ̣  ̣  ̣  ̣  ̣  ̣  ̣  ̣
            ευ  ̣  ̣  ̣  ̣ν τὸ ϲυντυγ-
            χάνειν. τοιούτων
            δὲ τῶν ἀγαθῶν ὄν-
 5         των ἡμῖν, καὶ ἐκ
            τῶν ἐναντίων
            τῶν κακῶν οὔτε
            ἀγαθόν τι ἔϲται – οὐ
            καλὸν δέ – οὔτε κα-
10        κόν – οὐκ αἰϲχρόν – οὔ-
            τε τὸ εὐδαιμονεῖν
\end{transcription}

{\footnotesize 2 \textgreek{εὔλ̣ο̣γ̣ο̣ν}? 11 perhaps \textgreek{τὸ εὐδαιμονεῖν || [ἐπιταθήϲεται ἢ ἀνεθήϲεται} Hammerstaedt}
\transversion
« … to happen. And such being the goods for us, even from the things that are opposed to evils there will be neither anything good–nor beautiful–nor anything bad–nor ugly–, nor happiness … »
\end{transpair}
\par

\noindent\textbf{Col.~21}\par\nobreak\vspace{\baselineskip}
\begin{transpair}
\begin{transcription}
            \emph{lost margin}
             ̣[  ̣  ̣  ̣  ̣]  ̣  ̣  ̣[  ̣ φ]ρο-
            νοῦν̣τεϲ μέγα καὶ
            ϲεμνυνόμενοι καὶ
            ll. 4-5 \emph{traces}
            ν̣οι  ̣  ̣  ̣  ̣  ̣  ̣τε ἀπὸ
            τοῦ α̣ὐτοῦ̣   ̣  ̣  ̣  ̣ ἐ̣παι̣-
            νε̣ῖ̣ν τ  ̣  ̣  ̣  ̣  ̣  ̣  ̣ν̣
            l. 9 \emph{traces}
10        μ̣ένου οἷον κατὰ τὰ̣ϲ̣
            ὑμνήϲειϲ εὐτ̣ον  ̣  ̣  ̣ 
\end{transcription}
{\footnotesize 11 \textgreek{εὐτ̣ονί̣α̣ι̣}?}
\transversion
« … being arrogant and feeling haughty and … from the same ... to praise … as according to the encomia … »
\end{transpair}
\par

\noindent\textbf{Col.~22}\par\nobreak\vspace{\baselineskip}
\begin{transpair}
\begin{transcription}
            \emph{lost margin}
              ̣ων ε̣ἰ̣ϲέτι ϲυνα̣  ̣  ̣  ̣  ̣[  ̣  ̣]
              ̣  ̣ Ἀριϲτοκ̣ρέω̣ν τα̣  ̣  ̣  ̣  ̣
              ̣  ̣  ̣(  ̣) τ̣οῖϲ ἐ̣χομένοιϲ̣  ̣  ̣
\end{transcription}

{\footnotesize 1 \textgreek{ϲυνα̣γ̣} or \textgreek{ϲυνα̣π̣} 2-3 perhaps \textgreek{τὰ̣ϲ̣ β̣ί|β̣λ̣ο̣υ̣ϲ̣}?\ Janko 3-4 \textgreek{β̣ι̣|[βλίοιϲ]}? Blank}
\transversion
« … still … Aristocreon [\emph{nominative/vocative}] … to the following … »
\end{transpair}
\flushbottom

%% file: sections/05c-methods-reporting.tex
\subsection*{Statistics and reproducibility}

The complete virtual unwrapping of \textit{PHerc.}~1667 amounted to 31~wraps and 1231~cm$^{2}$ of papyrus surface. All reported surface areas were calculated on the virtually unwrapped meshes. The unwrapping was completed using a semi-automated wrap by wrap copy tool combined with \textasciitilde{}25 hours per wrap of manual annotation.

The results comprise scroll-specific case studies of complete virtual unwrapping, direct volumetric ink visibility and title identification. The \textit{PHerc.}~1667 label-coverage ablation compares matched models qualitatively on an unlabeled held-out segment. Supplementary Tables~\ref{tab:sup1}--\ref{tab:sup4} report model configurations and, where available, the segmentation metrics and training or validation losses used to assess the models. Papyrological validation is based on review of the image evidence, geometric context and repeated renderings, with independent endorsement or consensus.

\subsection*{Use of AI tools}

AI tools assisted with language editing, structural revision, consolidation of cited background literature, figure assembly code and explanatory vector graphics. All scientific claims, interpretations, methodological descriptions and final wording were reviewed and approved by the human authors. The human authors validated the figure assets against their identified sources. Generative image tools were not used to alter empirical CT or ink images. No AI tools were used for reading, interpreting, and translating any Greek text. No large language model is listed as an author.

%% file: sections/07-backmatter.tex
\section*{Acknowledgements}

We thank the Director of the National Library of Naples “Vittorio Emanuele III”, Silvia Scipioni, and the curators, conservators, and staff members of the Library for access and guidance, in particular Giovanni Bova and Valentina De Martino. We acknowledge the European Synchrotron Radiation Facility (ESRF) for the provision of synchrotron radiation facilities on beamline BM18. 
The preliminary transcription of \textit{PHerc.}~1667 presented in this paper was first discussed at the conference The Eternal Library (Naples, 22--26 June 2026). We are grateful to the scholars who took part in the discussion for their valuable comments and suggestions, some of which we have incorporated into the present paper. These contributions are duly acknowledged in the critical apparatus in § “Complete \textit{PHerc.}~1667 transcription and translation” in Stage 5 (“Papyrological interpretation”).

\section*{Funding}

This research was supported by the Vesuvius Challenge (\url{https://scrollprize.org}), a non-profit donation-funded organization that aims to read the full collection of sealed scrolls from Herculaneum.

The Mellon Foundation sponsored the previous work by the EduceLab, University of Kentucky, that laid the groundwork for the realization of this research. 

\section*{Author contributions}

\begin{itemize}
\item G.A. led the project since June 2025, conceived the improved CT scanning protocol, contributed to photogrammetric acquisition, case design, ink segmentation models, mesh parametrization algorithms, scale-up on cloud architectures and drafted the first and the final versions of the manuscript.

\item S.P. led the project from February 2024 till June 2025, pioneered ink detection pipelines and contributed to photogrammetric acquisition, design of cases and scan protocol research.

\item F.N. led the papyrological efforts, selected the papyri from the National Library of Naples, and contributed to the photogrammetric acquisition and multispectral and infrared imaging.

\item Y.N. contributed to the ink detection models and pseudo-labeling.

\item S.J. developed the mesh tracing software and contributed to mesh tracing algorithms, ink detection models and annotation.

\item D.J. contributed to the annotation efforts since May 2023, led the annotation efforts since February 2025, and developed the whole-scroll banner merging algorithm.

\item P.H. supervised the machine learning and geometry research agenda and contributed to mesh fitting algorithms.

\item H.S. conceived and developed the mesh tracing software and contributed to mesh tracing algorithms.

\item J.R. conceived scaling on cloud architectures and contributed to data pipelines and software optimization.

\item F.M. contributed to the mesh tracing software.

\item E.R.D.P. contributed to fibers and ink detection models and annotation.

\item P.T. and A.M. supervised, conceived and implemented the CT data acquisition and reconstruction at the BM18 beamline.

\item C.S.P. pioneered CT scanning protocols and mesh tracing algorithms and contributed to the photogrammetry acquisition.

\item J.P.P. led the Vesuvius Challenge online contest from January 2023 till February 2024.

\item B.K. led the annotation efforts from May 2023 till February 2025.

\item A.L. contributed to papyrological review with particular focus on custodial and unrolling history of \textit{PHerc.}~1667, manuscript annotation, photogrammetric acquisition, and multispectral and infrared imaging.

\item R.V. contributed to papyrological review with particular focus on script and dating of \textit{PHerc.}~1667, manuscript annotation, photogrammetric acquisition, and multispectral and infrared imaging.

\item C.V. contributed to papyrological review with particular focus on \textit{PHerc.}~1667 coll. 1-12, manuscript annotation, photogrammetric acquisition, and multispectral and infrared imaging.

\item M.C.R. contributed to papyrological review with particular focus on \textit{PHerc.}~1667 coll. 13-22, manuscript annotation, photogrammetric acquisition, and multispectral and infrared imaging.

\item M.D.A. contributed to papyrological review with particular focus on \textit{PHerc.}~139,  manuscript annotation, photogrammetric acquisition, and multispectral and infrared imaging.

\item K.F. and M.McO. contributed to papyrological review and manuscript annotation.

\item G.D.M. performed papyrological review.

\item C.C. edited and rewrote sections of the paper, coordinated institutional relationships, permissions, conservator interactions and partner communication.

\item N.F. contributed to project conception, secured resources, advised on open-science and data-release strategy.

\item W.B.S. edited and rewrote sections of the paper, as well as initiated and pioneered the project and provided overall strategic guidance.
\end{itemize}

G.A., S.P., F.N., Y.N., S.J., D.J., P.H., H.S., J.R., F.M., E.R.D.P., P.T., A.M., J.P.P., C.V., M.D.A., G.D.M., W.B.S. acquired data on the BM18 beamline. G.A. and E.R.D.P. acquired data on the ID11 beamline.

\vspace{-7pt}
\section*{Competing interests}

The Authors declare no competing interests.

\vspace{-7pt}
\section*{Additional information}

Correspondence and requests for materials should be addressed to Giorgio Angelotti. W.~Brent Seales may also be contacted for material and general inquiries.

\vspace{-7pt}
\section*{Data availability}\label{sec:dataavailability}

Tomographic volumes used for all analyses in this work (unsigned 8-bit, masked OME-Zarr; see Methods), reconstructed surfaces, flattened renderings, figure source data and transcription artifacts are available at \url{https://scrollprize.org/data_browser}. Data is shared under the Creative Commons 4.0 - BY NC License unless otherwise specified. JPEG2000 stack versions of the original 16-bit ESRF reconstructions, prior to 8-bit conversion, masking and OME-Zarr export, are also available on the ESRF open access database for art, history and archaeology\cite{dataESRF1667,dataESRF139,dataESRFParis4} under the same license.

\vspace{-7pt}
\section*{Code availability}

Code sufficient to reproduce the main analyses, including mesh tracing, flattening, surface-conditioned rendering and ink-detection inference, is available at \url{https://github.com/ScrollPrize/villa/commit/e583fb67468f483fadd73d52e06f0ab0fe5ba813}.

ML model training configurations and datasets are listed in Supplementary Tables~\ref{tab:sup1}--\ref{tab:sup4}; model weights are available on the Vesuvius Challenge Hugging Face organization (\url{https://huggingface.co/scrollprize}). The \textit{PHerc.}~1667 ink-detection ablation checkpoints are stored at \url{https://huggingface.co/collections/scrollprize/pherc1667-ink-detection-ablation}.

%% file: sections/08-extended-data.tex
\phantomsection
\section*{Extended Data}\label{sec:extdata}

\renewcommand{\figurename}{Extended Data Fig.}
\renewcommand{\tablename}{Extended Data Table}
\setcounter{figure}{0}
\setcounter{table}{0}

\begin{sidewaystable}
  \centering\footnotesize
  \caption{\textbf{Per-scroll summary of scan parameters.}}
  \label{tab:ed1}
  \begin{tabular}{@{}l >{\RaggedRight\arraybackslash}p{32mm} l c c c c c c@{}}
    \toprule
    Sample ID & Role in manuscript & Beamline & Avg.\ energy (keV) & Voxel / pixel size (µm) & Sample-to-detector distance (m) & Phase retrieval δ/β & Unsharp coeff. & \makecell[b]{Unsharp σ\\(pixels)} \\
    \midrule
    \textit{PHerc.}~9B fragment    & Fragment-derived ink supervision                                   & BM18 & 77  & 2.401 & 0.35 & 1000 & 4 & 1.2 \\
    \textit{PHerc.}~343P fragment  & Fragment-derived ink supervision; scan-calibration reference      & BM18 & 77  & 2.403 & 0.22 & 1000 & 4 & 1.2 \\
    \textit{PHerc.}~500P2 fragment & Fragment-derived ink supervision                                   & BM18 & 111 & 2.215 & 0.40 & 1000 & 4 & 1.2 \\
    \textit{PHerc.}~500P2 fragment & High-resolution fragment comparison                                & ID11 & 65  & 0.55  & 0.07 & 20   & — & — \\
    \textit{PHerc.~Paris.}~4        & Volumetric ink-validation case                                     & BM18 & 78  & 2.400 & 0.22 & 1000 & 4 & 1.2 \\
    \textit{PHerc.~Paris.}~3        & Scanned comparison volume                                          & BM18 & 78  & 2.400 & 0.22 & 1000 & 4 & 1.2 \\
    \textit{PHerc.}~1667           & Principal complete-unwrapping result                               & BM18 & 78  & 2.399 & 0.22 & 1000 & 4 & 1.2 \\
    \textit{PHerc.}~1667 (slab)    & Complement reading, detector-pixel-size comparison                 & BM18 & 59  & 1.129 & 0.22 & 500  & 4 & 1.2 \\
    \textit{PHerc.}~1451           & Scanned comparison / training volume                               & BM18 & 78  & 2.399 & 0.22 & 1000 & 4 & 1.2 \\
    \textit{PHerc.}~332            & Scanned comparison / training volume                               & BM18 & 78  & 2.399 & 0.22 & 1000 & 4 & 1.2 \\
    \textit{PHerc.}~139            & Title-identification result                                        & BM18 & 78  & 2.399 & 0.22 & 1000 & 4 & 1.2 \\
    \textit{PHerc.}~814            & Scanned comparison / training volume                               & BM18 & 78  & 2.399 & 0.22 & 1000 & 4 & 1.2 \\
    \textit{PHerc.}~841 slab       & Slab scan / comparison volume                                      & BM18 & 77  & 2.403 & 0.22 & 1000 & 4 & 1.2 \\
    \textit{PHerc.}~1203 slab      & Slab scan / comparison volume                                      & BM18 & 77  & 2.403 & 0.22 & 1000 & 4 & 1.2 \\
    \textit{PHerc.}~1299           & Scanned comparison / training volume                               & BM18 & 78  & 2.399 & 0.22 & 1000 & 4 & 1.2 \\
    MANB                   & Scanned comparison / training volume                               & BM18 & 78  & 2.399 & 0.22 & 1000 & 4 & 1.2 \\
    MANBp fragment         & Fragment / training volume                                         & BM18 & 78  & 2.399 & 0.22 & 1000 & 4 & 1.2 \\
    MAN5                   & Scanned comparison / training volume                               & BM18 & 78  & 2.399 & 0.22 & 1000 & 4 & 1.2 \\
    \bottomrule
  \end{tabular}
\end{sidewaystable}

\begin{figure}[H]
  \centering
  \includegraphics[width=\linewidth]{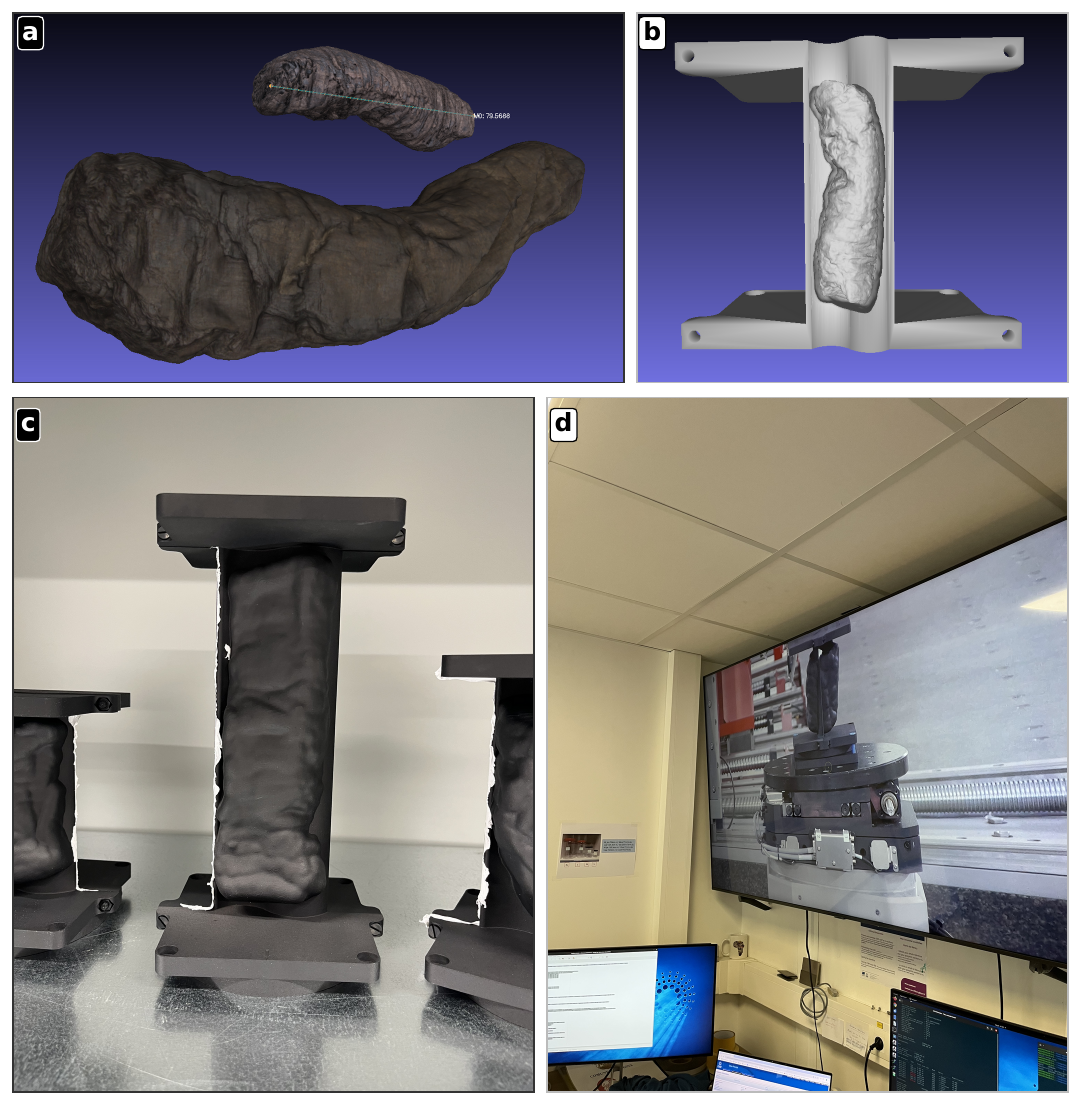}
  \caption{\textbf{Sample preparation and scan cases.} \textbf{a}, Photogrammetric meshes of \textit{PHerc.}~1667 (top) and \textit{PHerc.~Paris.}~4 (bottom), reconstructed from multi-view photographs; the M0 label gives the maximum chord length of \textit{PHerc.}~1667 in millimeters. \textbf{b}, Automatically generated half-case model for \textit{PHerc.}~1667 (non-textured render), with the photogrammetric scroll silhouette visible inside the cavity; the case is fitted to the outer morphology to minimize scan time, dose exposure and movement. \textbf{c}, Multi-jet-fusion-printed nylon cases vertically standing with scrolls inside. \textbf{d}, A cased sample mounted on the beamline rotation stage, viewed on the operator's monitor during acquisition. The scrolls shown in c and d are illustrative and are not among the objects reported in this paper; see \hyperref[sec:methods]{Methods} for case generation, materials and acquisition parameters.}
  \label{fig:ed1}
\end{figure}

\clearpage
\begingroup\centering
  \captionof{figure}{\textbf{Imaging trade-offs in densely packed Herculaneum papyrus.} Representative reconstructions showing the effects of pixel size, propagation distance, incident energy and phase retrieval on layer separability and haze-like fiber effects.}
  \label{fig:ed2}\par\medskip

\includegraphics[width=\linewidth]{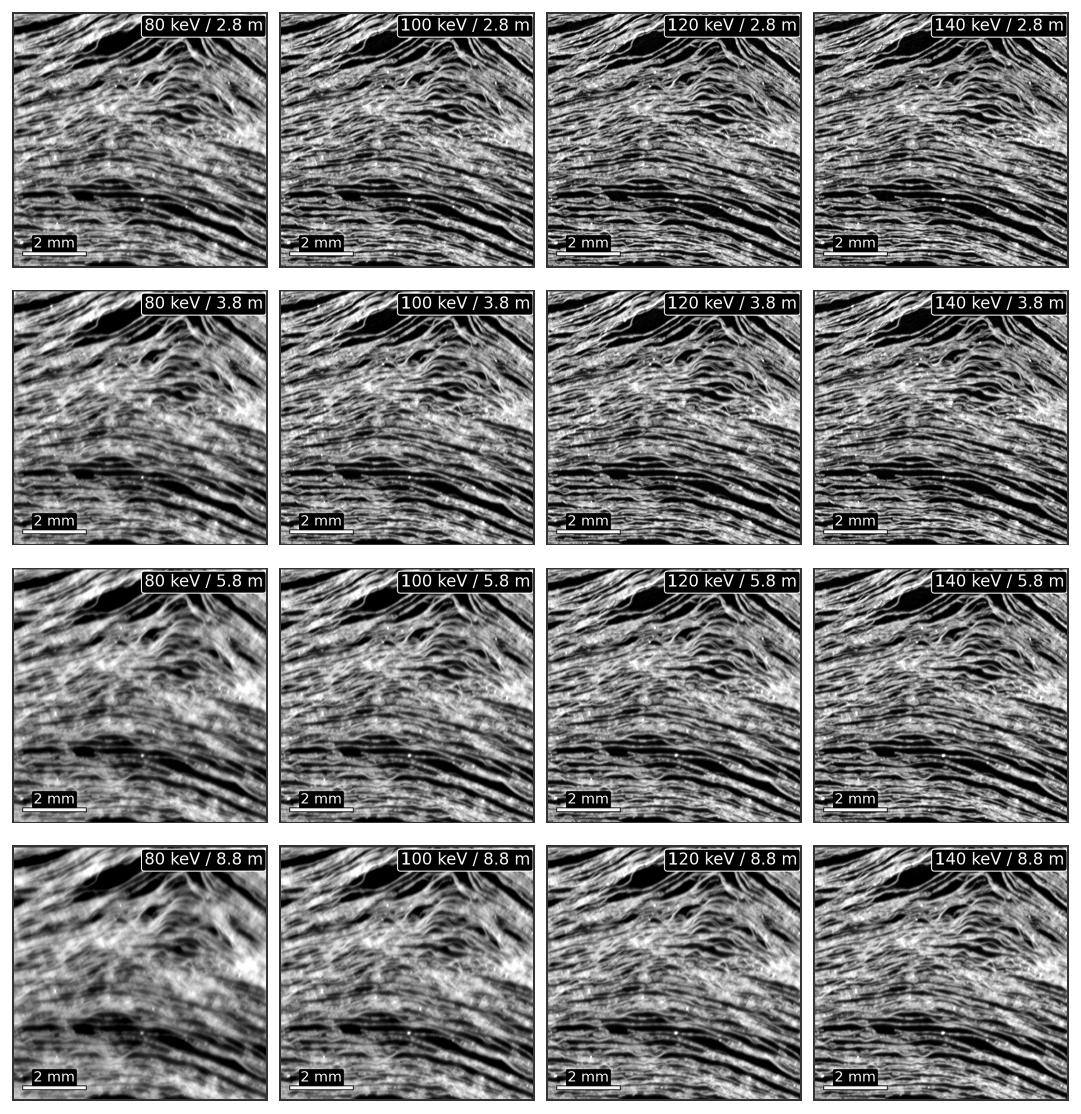}\par
{\small\textbf{a}, \textbf{Joint propagation-distance × incident-energy sweep at fixed pixel size.} Sixteen axial reconstructions of the same \textit{PHerc.}~175A slice in a 4 × 4 matrix: rows (top → bottom) are sample-to-detector propagation distance 2.8 / 3.8 / 5.8 / 8.8 m; columns (left → right) are mean incident energy 80 / 100 / 120 / 140~keV. All cells share detector pixel size (8.009–8.013~µm effective), Paganin filter (δ/β $=$ 1000). Physical field of view 8~mm × 8~mm per cell; scale bar 2~mm. With pixel size held constant, reading rows top-to-bottom (fixed energy, increasing distance) shows the bright/dark phase-fringe halos broadening around every layer interface; reading columns left-to-right (fixed distance, increasing energy) shows the absorption contrast flattening as the mean spectrum-energy rises. These tests show the decoherence effect that appears as a blurring of the slices. The most visible case is for the lower left corner, that corresponds to the lowest energy combined with the longest distance. The decoherence effect then decreases if the energy increases, and/or if the propagation distance decreases. The operationally usable corner of the matrix sits above the top-left to bottom-right diagonal, with image quality increasing with decreasing sample-to-detector propagation distance. For a pixel size of about 8~µm the empiric sweet spot for the energy is between 100 and 120~keV with propagation distances of about 3~m.}\par\medskip

\includegraphics[width=\linewidth]{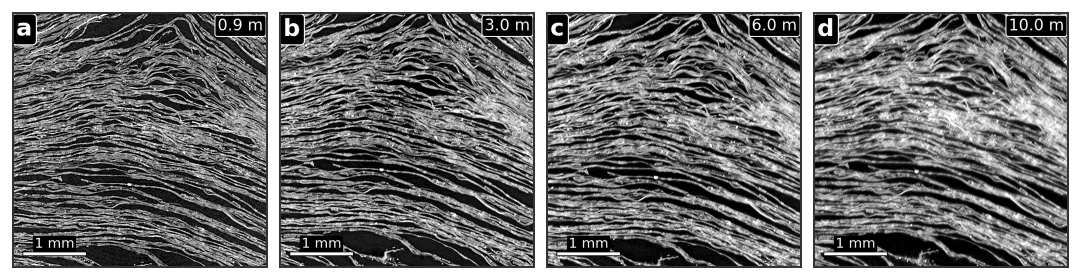}\par
{\small\textbf{b}, \textbf{Sample-to-detector propagation distance effect on retrieved phase-contrast at approximately fixed pixel size and energy.} Four reconstructions of the same \textit{PHerc.}~175A slice at fixed mean energy (133~keV), fixed Paganin filter (δ/β $=$ 1000) and effective pixel size 4.1--4.3~µm as detector magnification varies with distance: 0.9 m; 3.0 m; 6.0 m; 10.0 m. Physical field of view 4~mm × 4~mm. Image quality increases with decreasing propagation distance due to the progressively increasing decoherence effect with increasing distance.}\par\medskip

\includegraphics[width=\linewidth]{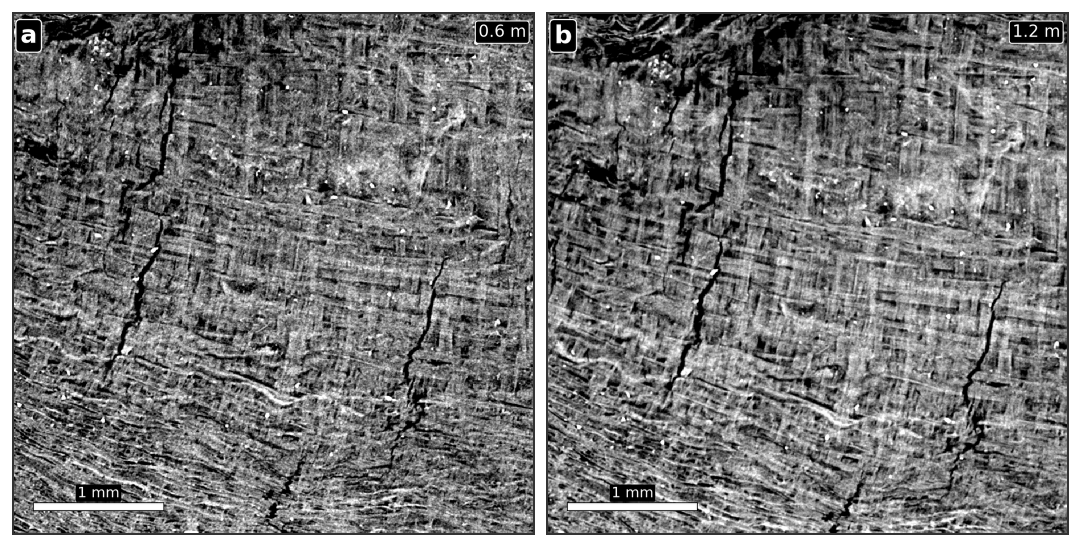}\par
{\small\textbf{c}, \textbf{Sample-to-detector propagation distance at fixed pixel size and energy.} Two reconstructions of slices from \textit{PHerc.}~268 at fixed mean energy (133~keV), fixed Paganin filter (δ/β $=$ 1000), and effectively matched pixel size (4.334~µm at 0.6 m, 4.320~µm at 1.2 m): 0.6 m; 1.2 m. Physical field of view 4~mm × 4~mm; scale bar 1~mm. At 0.6 m, the crack in the slice reads as a thin dark line and the surrounding layers are well defined. Even if these slices look similar to absorption contrast, it is directly due to the phase retrieval process. In fact, even at the shortest propagation distances, the contrast is strongly dominated by phase-contrast. Absorption contrast remains very low as the samples are mostly composed of carbon that absorbs faintly in the hard X-ray regime. The picture on the right gives an overall better visibility of the cracks and of some other structures, but at the expense of the sharpness in many places. Once again, this is due to the decoherence effect that starts to affect the resolution at 1.2~m. This comparison shows that the phase-contrast is necessary to produce usable level of contrast in these samples, but the propagation distance has to be kept low enough not to reach the decoherence level that would impact the resolution. This distance can change from one sample to another one depending on its diameter and packing density. The denser and larger it is, and the higher the decoherence effect is. These tests informed the selected propagation distance of 220~mm for the 2.4~µm pixel size setup, balancing contrast against decoherence-related blurring in large scrolls. This distance can be longer when larger pixel sizes are used (about 0.5~m at 4.6~µm and 2~m at 8~µm).  \textit{PHerc.}~268 is the only sample in the campaign with a single-sample matched sub-1 m / above-1 m pair; the full 0.9 / 3 / 6 / 10 m sweep in the previous image is on \textit{PHerc.}~175A.}\par\medskip

\includegraphics[width=\linewidth]{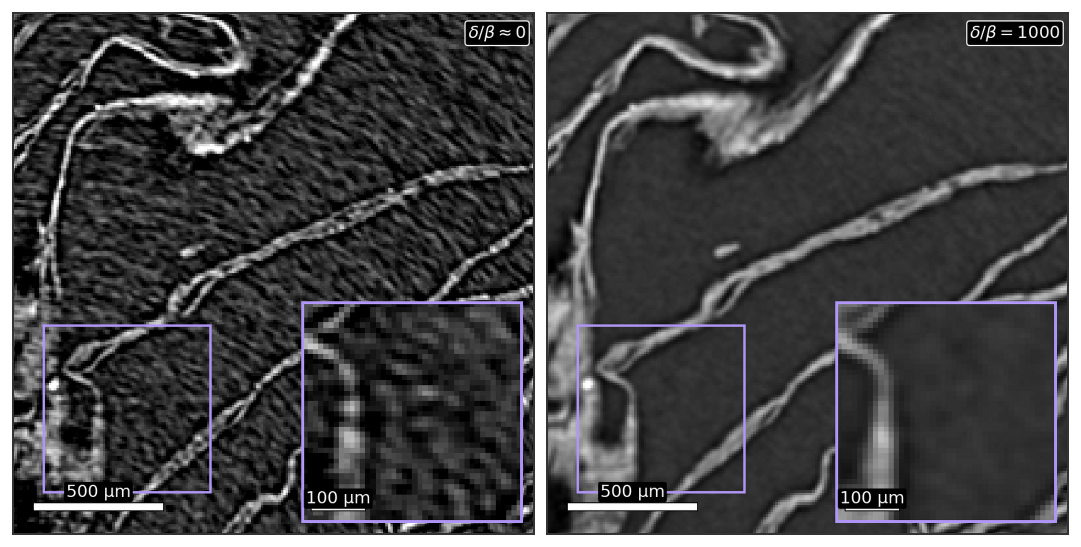}\par
{\small\textbf{d}, \textbf{Companion view of the Paganin phase-retrieval trade-off on \textit{PHerc.}~139 ( 9.362~µm pixel size).} \textit{PHerc.}~139, 113~keV mean, 1.2 m sample-to-detector, 9.362~µm effective pixel size, binmean2 of a 4.681~µm acquisition processed with two Paganin δ/β values illustrate the phase-retrieval trade-off. From left: δ/β ≈ 0 (no phase retrieval, equivalent to the raw filtered) back-projection and δ/β $=$ 1000 (the production setting used for the principal 2.4~µm unwrapping and ink-detection scans). Each panel covers a 2.0~mm × 2.0~mm field of view at native pyramid level 0; the violet rectangle marks a 400 × 400~µm sub-ROI and the  bottom-right inset is that ROI at 5× magnification. Scale bars are 500~µm (main) and 100~µm (zoom). At δ/β ≈ 0 fringes are visible, the image is more noisy but has also higher frequency texture signal; δ/β $=$ 1000 (production) suppresses the fringes and obtains higher SNR and contrast at the cost of reduced apparent sharpness.}\par\medskip

\includegraphics[width=\linewidth]{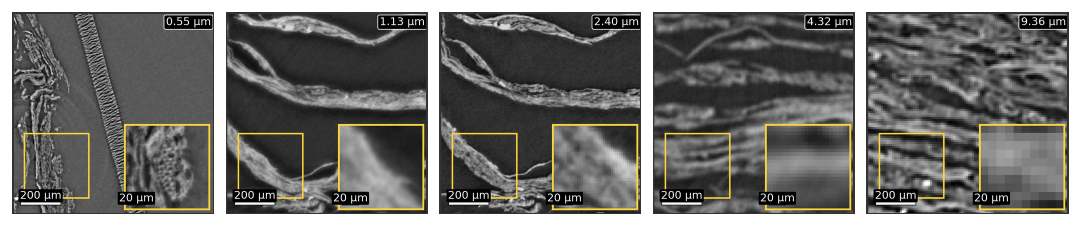}\par
{\small\textbf{e}, \textbf{Layer visibility across detector pixel sizes and acquisition conditions.} Reconstructions of densely packed Herculaneum papyrus illustrate layer visibility at five detector pixel sizes. Specimen, energy, propagation distance and phase-retrieval settings also differ, so this comparison does not isolate the effect of pixel size. From left: 0.55~µm (\textit{PHerc.}~500P2 ID11 monochromatic small-ROI scan, 65~keV, 0.07 m propagation,  δ/β $=$ 20), 1.129~µm (\textit{PHerc.}~1667, BM18 59~keV polychromatic, 0.22~m propagation), 2.399~µm (\textit{PHerc.}~1667, BM18 78~keV polychromatic, 0.22 m propagation), 4.317~µm (\textit{PHerc.}~500P2, BM18 111~keV polychromatic, 1.2 m propagation), and 9.362~µm effective (\textit{PHerc.}~500P2 fragment HA, BM18 113~keV polychromatic, 1.2 m, exported with binmean2 detector binning so the effective pixel size is twice). Each panel covers a 1.0~mm × 1.0~mm axial field of view at its native pyramid level 0; the yellow rectangle marks a 100 × 100~µm sub-ROI and the bottom-right inset is that ROI at 10× magnification. Scale bars at the bottom of each panel are 200~µm (main) and 20~µm (zoom inset). At 0.55~µm individual papyrus fibers are resolved (cellular cross-sections visible in the inset); at 1.129~µm fine delaminations begin to blur due to a too long propagation distance for a so small pixel size.; 2.4~µm was the selected sampling for complete imaging of large scrolls;  the 4.317~µm panel is haze-limited and the 9.362~µm panel is pixel-limited, the inset shows the bare voxel grid.}\par\medskip

\includegraphics[width=\linewidth]{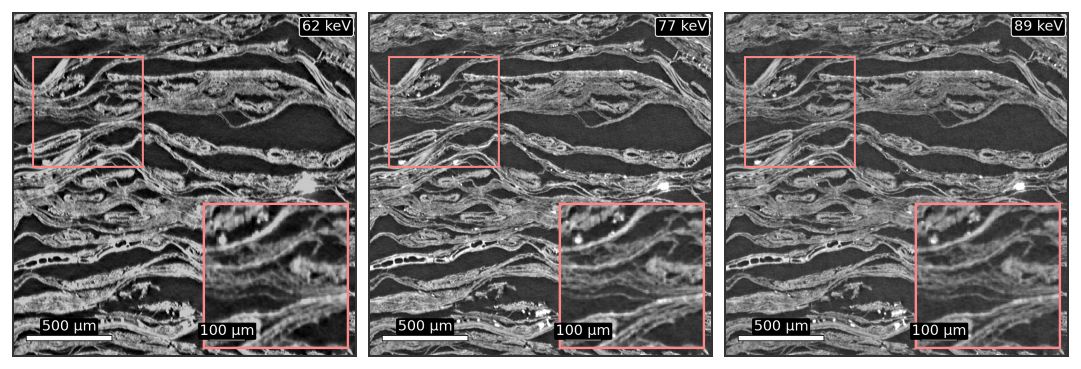}\par
{\small\textbf{f}, \textbf{Incident X-ray energy trades contrast magnitude and beam decoherence.} Three slice reconstructions of \textit{PHerc.}~343P 2.4~µm / 0.22 m / δ/β $=$ 1000, 62~keV; 77~keV; 89~keV. Physical size 2.0~mm × 2.0~mm FOV. Across 62 → 89~keV, the decrease in overall contrast (faint absorption with superimposed propagation phase-contrast) is visible but gradual. Layer boundaries remain clear in every column, while layer separability increases. We found the empirical sweet spot to be at 77~keV.}\par\medskip

\par
\endgroup
\clearpage

\begin{figure}[H]
  \centering
  \includegraphics[width=\linewidth]{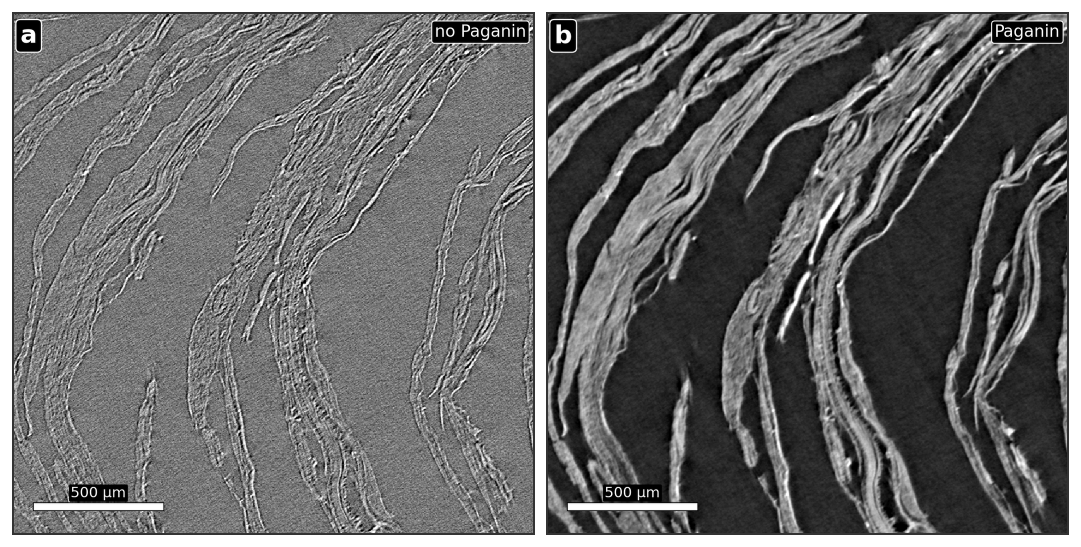}
  \caption{\textbf{Ink directly visible in \textit{PHerc.~Paris.}~4.} Comparison of the same region of interest in the \textit{PHerc.~Paris.}~4, 78~keV, 2.4~µm pixel size, propagation distance 0.22 m \textbf{a}, No Paganin (or more accurately δ/β $=$ 0.0001). The slice is dominated by phase contrast fringes. Contrast is weak and washed out. \textbf{b}, Paganin (δ/β $=$ 1000). Contrast is strong, separation of different structures becomes clearer, ink flakes (in the center) are brighter than the surrounding papyrus.}
  \label{fig:ed3}
\end{figure}

\begin{figure}[p]
  \centering

  \includegraphics[width=0.8\linewidth]{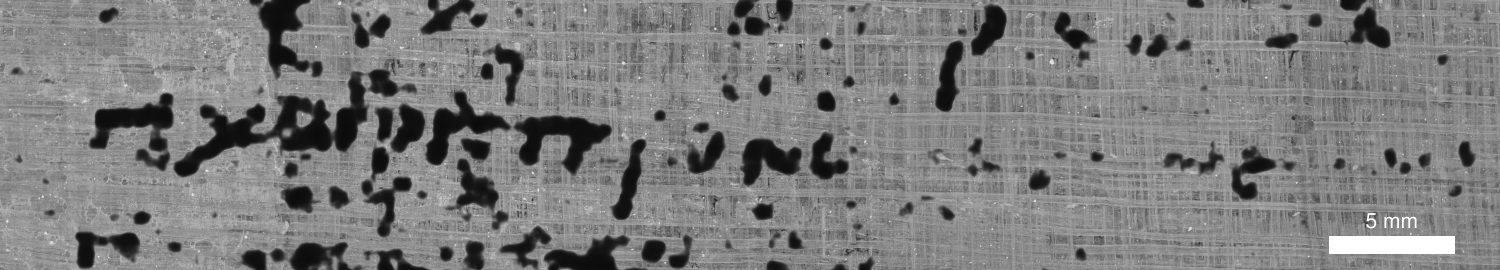}\\[2pt]
  \textbf{a}\quad w49: \textgreek{χωρὶϲ προνο[ί]α̣ϲ̣} (“without providence”)

  \vspace{0.9em}
  \includegraphics[width=0.8\linewidth]{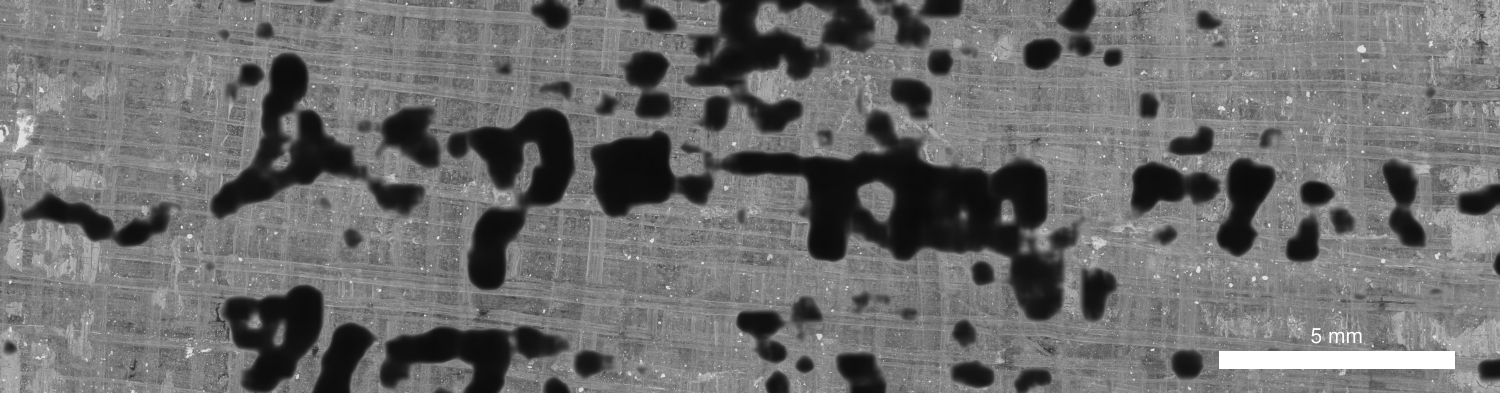}\\[2pt]
  \textbf{b}\quad w25: \textgreek{ἀόρατα} (“invisible entities”)

  \vspace{0.9em}
  \includegraphics[width=0.8\linewidth]{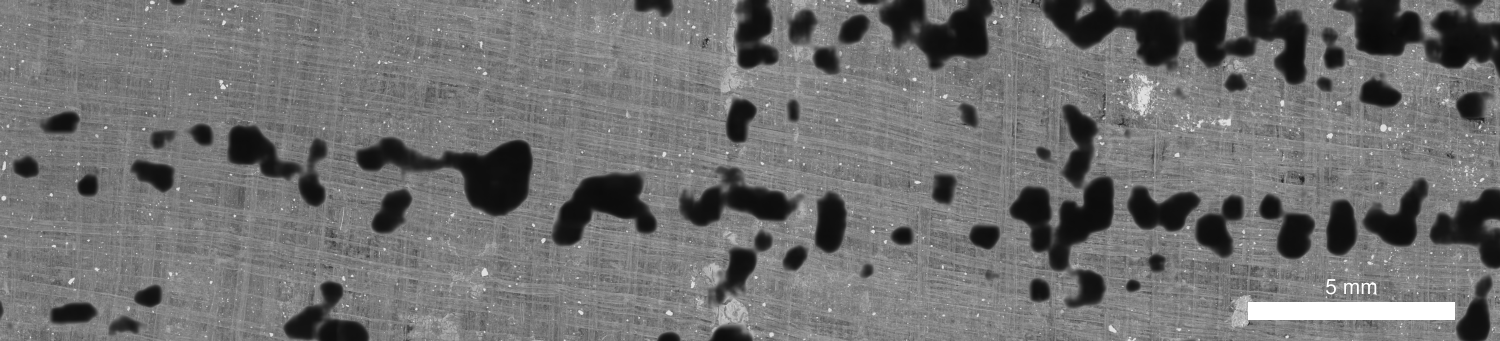}\\[2pt]
  \textbf{c}\quad w34: \textgreek{καὶ τὸ κατ̣ὰ̣ φύϲιν} (“and the \ldots{} according to nature”)

  \vspace{0.9em}
  \includegraphics[width=0.8\linewidth]{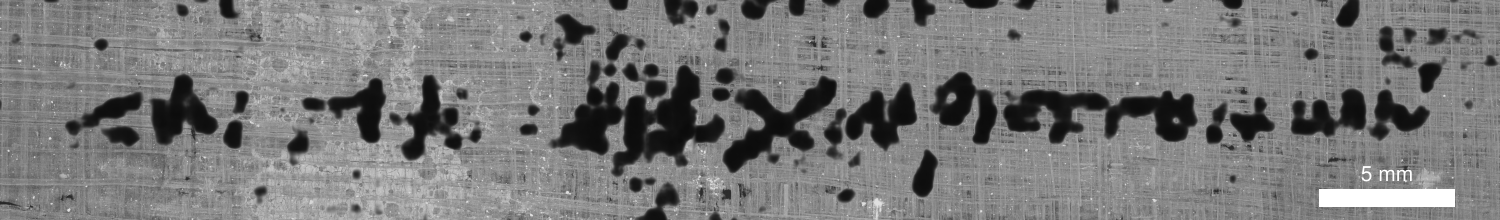}\\[2pt]
  \textbf{d}\quad w47: \textgreek{καὶ πάν̣τ̣α̣ χωρὶϲ πόνων} (“and everything without troubles”)

  \vspace{0.9em}
  \includegraphics[width=0.8\linewidth]{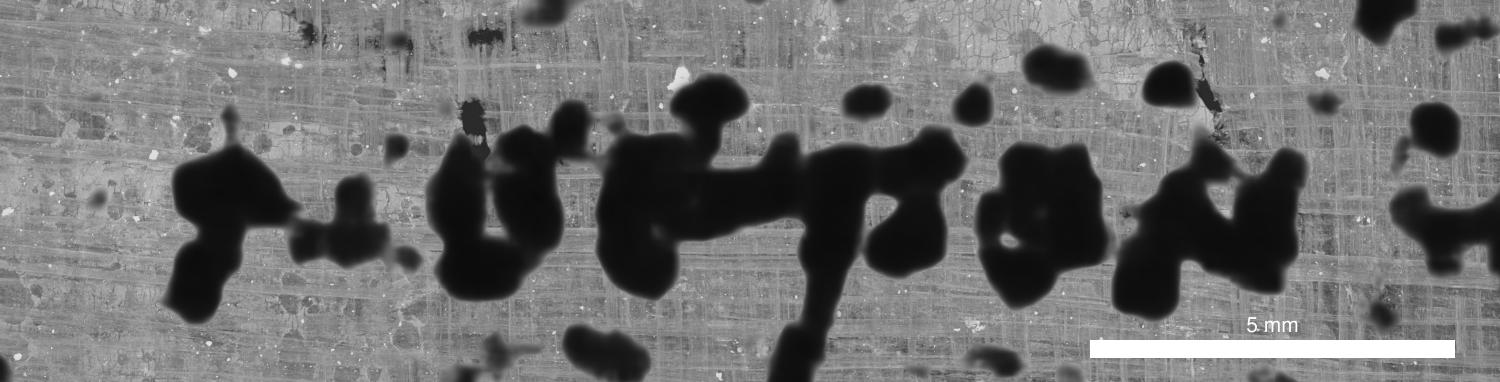}\\[2pt]
  \textbf{e}\quad w49: \textgreek{νοερόν} (“intellectual”)

  \caption{\textbf{Readings from \textit{PHerc.}~139.} Ink predictions are shown in black over maximum-projection CT composites. Scale bars, 5~mm.}
  \label{fig:sup1}

\end{figure}
\FloatBarrier %

\begin{figure}[H]
  \centering
  \includegraphics[width=\linewidth]{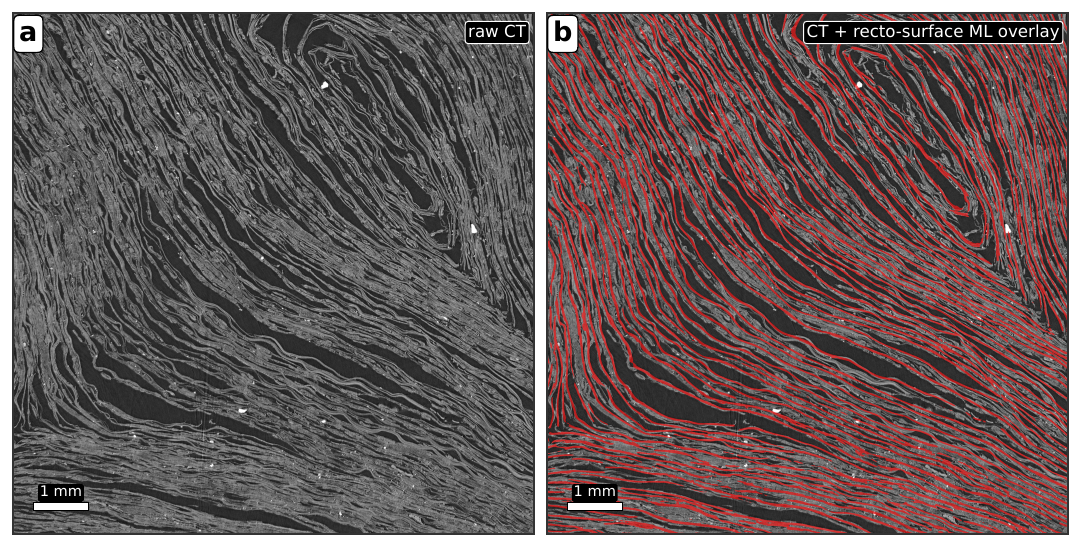}
  \caption{\textbf{Surface predictions.} Examples of recto-surface prediction in \textit{PHerc.~Paris.}~4.}
  \label{fig:ed4}
\end{figure}

\begin{figure}[H]
  \centering
  \includegraphics[width=\linewidth]{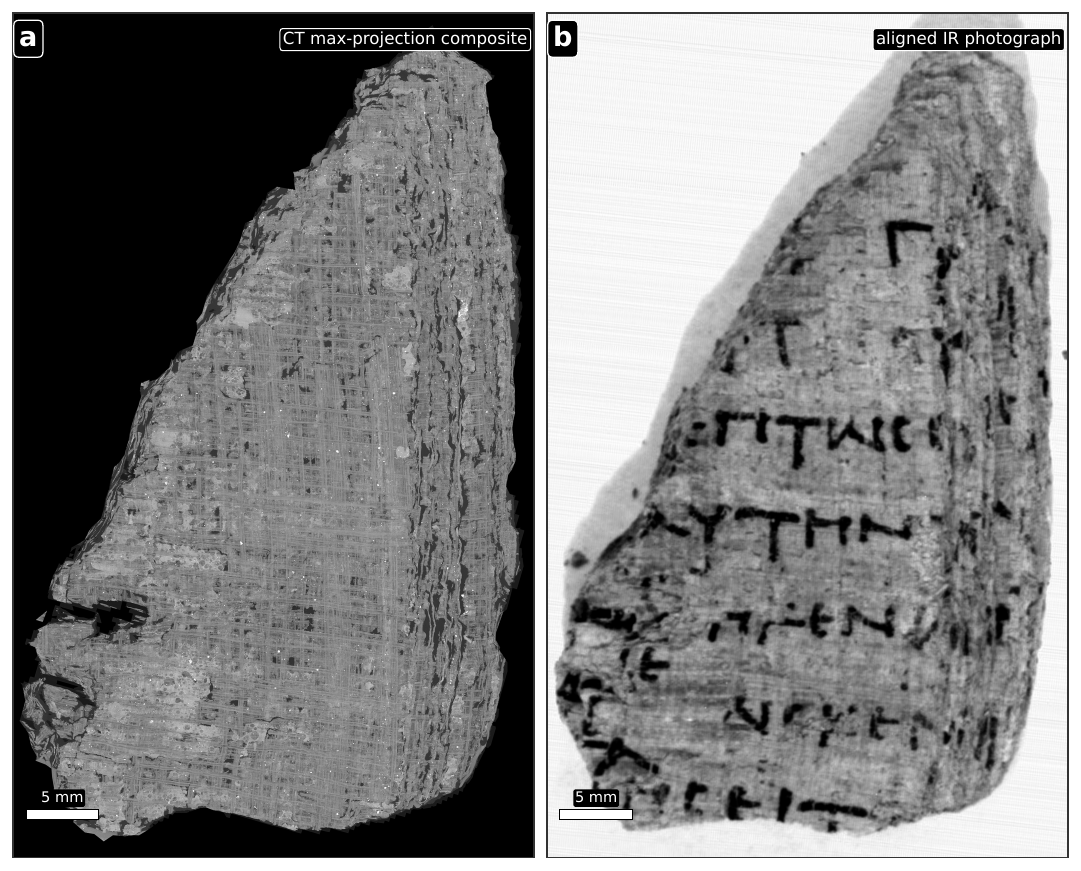}
  \caption{\textbf{Fragment-derived supervision.} \textit{PHerc.}~500P2 front-surface segment, a $=$ CT max-projection composite (BM18 2.215~µm / 111~keV / 0.4 m / Paganin δ/β $=$ 1000, orthorectified, integrating 65 ortho-layers), b $=$ Naples M1050 IR photograph \#017 warped to the composite's coordinates via manually-placed control points. 35.8 × 58.1~mm physical FOV. Scale bars 5~mm.}
  \label{fig:ed5}
\end{figure}

\begin{figure}[H]
  \centering
  \includegraphics[width=\linewidth]{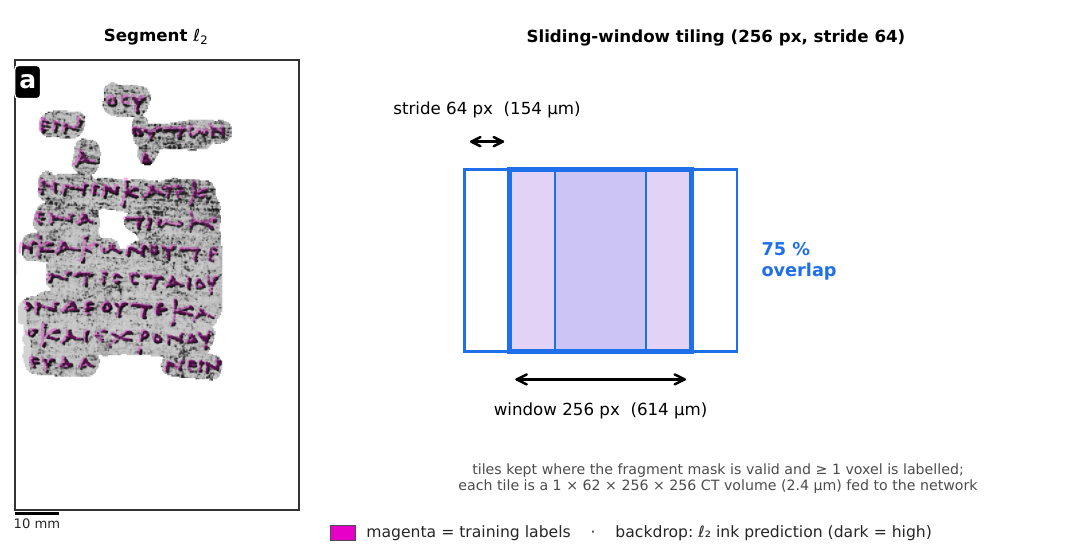}
  \caption{\textbf{Sliding-window sampling and model input geometry.} \textbf{a}, Ink predictions on \textit{PHerc.}~1667 segment $\ell_2$, with training labels in magenta. \textbf{b}, Schematic 256-pixel windows at 64-pixel stride (75\% overlap along each axis). Each detector input contains 62 × 256 × 256~voxels from the surface-aligned CT stack. Scale bar, 10~mm (a).}
  \label{fig:ed6}
\end{figure}

\begin{figure}[H]
  \centering
  \includegraphics[width=\linewidth]{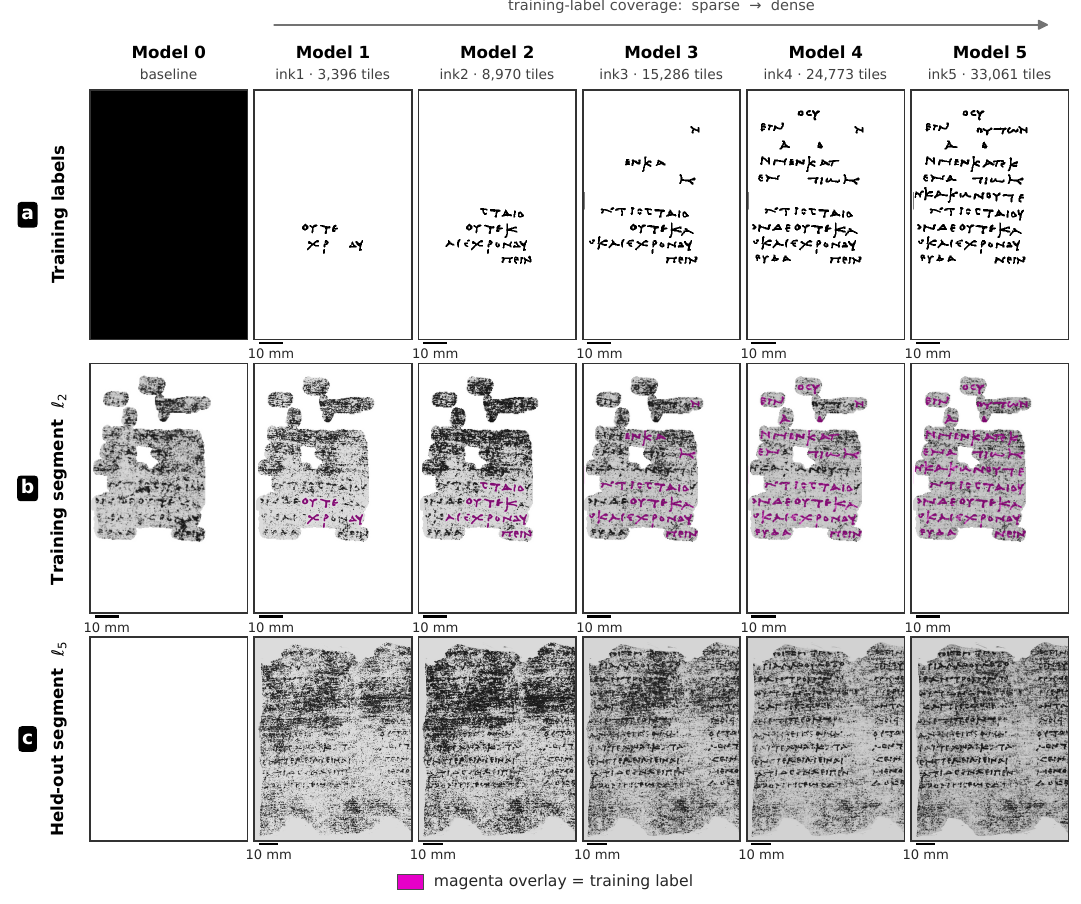}
  \caption{\textbf{Pseudo-labeling ablation.} Models 1--5 were trained from a common initialization pretrained on the combined Kinetics-700 and Moments in Time datasets with the same training-step count, using successively expanded label sets. Model 0 is the starting model, not trained on \textit{PHerc.}~1667. Row a shows the training labels. Row b shows predictions on the training segment, with labels overlaid in magenta. Row c shows predictions on the held-out segment, for which reference ink labels are unavailable.}
  \label{fig:ed7}
\end{figure}

%% file: sections/09-supplementary.tex
\renewcommand{\figurename}{Supplementary Fig.}
\renewcommand{\tablename}{Supplementary Table}
\setcounter{figure}{0}
\setcounter{table}{0}

\renewcommand{\textfraction}{0.0}
\renewcommand{\topfraction}{1.0}
\renewcommand{\bottomfraction}{1.0}
\renewcommand{\floatpagefraction}{0.5}
\setcounter{topnumber}{3}
\setcounter{totalnumber}{4}

\begin{sidewaystable}[p]
  \phantomsection\label{sec:supp}
  {\large\bfseries Supplementary Information\par}\medskip
  \centering\footnotesize\setlength{\tabcolsep}{3pt}
  \caption{\textbf{Scans used for the experimental campaign}}
  \label{tab:sup5}
  \begin{tabular}{@{}l >{\RaggedRight\arraybackslash}p{22mm} l l l l l >{\RaggedRight\arraybackslash}p{34mm}@{}}
    \toprule
    Sample ID & Figure / comparison & Beamline & \makecell[b]{Avg. energy\\(keV)} & \makecell[b]{Voxel / pixel\\size (µm)} & \makecell[b]{Sample-to-detector\\distance (m)} & \makecell[b]{Phase retrieval\\δ/β} & Reconstruction / processing note \\
    \midrule
    \textit{PHerc.}~175A & Energy × propagation-distance sweep & BM18 & 80, 100, 120, 140 & 8.009--8.013 & 2.8, 3.8, 5.8, 8.8 & 1000 & 4 × 4 matrix; 16 axial reconstructions of the same slice. \\
    \textit{PHerc.}~175A & Propagation-distance sweep at fixed energy & BM18 & 133 & 4.1--4.3 & 0.9, 3.0, 6.0, 10.0 & 1000 & Four reconstructions; effective pixel size varies slightly with detector magnification. \\
    \textit{PHerc.}~175A & Paganin phase-retrieval comparison & BM18 & 111 & 2.215 & 0.40 & 0 and 1000 & Same ROI reconstructed without phase retrieval and with production Paganin setting. \\
    \textit{PHerc.}~268 & Matched sub-1 m / above-1 m propagation comparison & BM18 & 133 & 4.334; 4.320 & 0.6; 1.2 & 1000 & Two matched reconstructions at similar effective pixel size. \\
    \textit{PHerc.}~1667 & Detector-pixel-size comparison & BM18 & 59 & 1.129 & 0.22 & 500 & Used as the 1.129 µm panel in the pixel-size comparison; unsharp coefficient 4 and σ = 1.2 pixels. \\
    \textit{PHerc.}~1667 & Detector-pixel-size comparison / production reference & BM18 & 78 & 2.399 & 0.22 & 1000 & Same as principal \textit{PHerc.}~1667 production scan; included here only as a comparison panel. \\
    \textit{PHerc.}~500P2 fragment & Detector-pixel-size comparison & ID11 & 65 & 0.55 & 0.07 & 20 & Monochromatic small-ROI scan; highest-resolution comparison panel. \\
    \textit{PHerc.}~500P2 fragment & Detector-pixel-size comparison & BM18 & 111 & 4.317 & 1.2 & 1000 & Used as the 4.317 µm panel. \\
    \textit{PHerc.}~500P2 fragment HA & Detector-pixel-size comparison & BM18 & 113 & 9.362 effective & 1.2 & 1000 & Exported with binmean2 from a 4.681 µm acquisition; coarsest pixel-size panel. \\
    \textit{PHerc.}~139 & Paganin phase-retrieval comparison & BM18 & 113 & 9.362 effective & 1.2 & ≈0 and 1000 & Same acquisition processed with no/near-no Paganin and with δ/β = 1000. \\
    \textit{PHerc.}~343P fragment & Incident-energy comparison & BM18 & 62, 77, 89 & 2.4 & 0.22 & 1000 & Three reconstructions used to compare energy-dependent contrast and layer separability. \\
    \textit{PHerc.}~343P fragment & Propagation-distance comparison & BM18 & 77 & 2.401--2.403 & 0.22; 0.35 & 1000 & Sequential acquisitions differing only in propagation distance. \\
    \textit{PHerc.~Paris.}~4 & No-Paganin / Paganin ink-visibility comparison & BM18 & 78 & 2.4 & 0.22 & 0.0001 and 1000 & Same ROI reconstructed with near-no Paganin and with δ/β = 1000. \\
    \textit{PHerc.~Paris.}~4 & Production volumetric ink-validation reconstruction & BM18 & 78 & 2.4 & 0.22 & 1000 & Same production volume as in Extended Data Table~\ref{tab:ed1}; included here because it is compared against the no-Paganin reconstruction. \\
    \bottomrule
  \end{tabular}
\end{sidewaystable}

\FloatBarrier

\begin{table}[htbp]
  \centering
  \footnotesize
  \caption{\textbf{Training configuration of the recto-surface 3D U-Net.} Values were taken from the Weights \& Biases run ps256\_bs2 (d5jdo9n1, 2026-03-31), archived in project vesuvius-challenge/paper.}
  \label{tab:sup1}
  \begin{tabularx}{\textwidth}{@{}l l X@{}}
    \toprule
    Category & Parameter & Value \\
    \midrule
    \multirow{4}{*}{Architecture}
      & Backbone & nnU-Net-style 3D residual-encoder U-Net \\
      & Convolution & nn.Conv3d \\
      & Normalization & nn.InstanceNorm3d \\
      & Attention & Concurrent spatial/channel squeeze-and-excitation (scSE) \\
    \cmidrule(lr){2-3}
    \multirow{4}{*}{Input}
      & Output channels & 2 (background / surface) \\
      & Patch size (Z × Y × X) & 256 × 256 × 256 voxels \\
      & Intensity normalization & Per-patch z-score \\
      & Ignore label & 2 (held-out / mask-out) \\
    \cmidrule(lr){2-3}
    \multirow{5}{*}{Training data}
      & Volumes & \textit{PHerc.}~0139 (three ledger pages of labels), \textit{PHerc.}~1667, \textit{PHerc.}~0343P, \textit{PHerc.}~0500P2, \textit{PHerc.}~MAN Bp \\
      & Train patches & 116,531 \\
      & Validation patches & 2,379 \\
      & Train / val split & 0.98 / 0.02 (seed = 42) \\
      & Min labeled-ratio gate & 0.001 \\
    \cmidrule(lr){2-3}
    \multirow{11}{*}{Optimization}
      & Min bbox-fill gate & 0.35 \\
      & Loss & Medial Surface Recall (custom Skeleton-Recall adaptation; combined with cross-entropy and soft Dice on the two-channel output, weight = 1) \\
      & Batch size & 2 \\
      & Mixed precision & AMP fp16 \\
      & Gradient clipping (L2 norm) & 12 \\
      & Validation cadence & every 5 epochs (≤ 100 validation steps/epoch) \\
      & Max epochs (scheduled) & 7,500 \\
      & Final logged epoch & 3,864 (run end-state: terminated) \\
    \cmidrule(lr){2-3}
    \multirow{4}{*}{\makecell[tl]{Final validation\\(epoch 3,864)}}
      & Learning rate at final epoch & $5.21\times10^{-3}$ \\
      & Mean Dice / IoU & 0.596 / 0.495 \\
      & Surface (class 1) Dice / IoU & 0.308 / 0.189 \\
      & Background (class 0) Dice / IoU & 0.883 / 0.801 \\
    \bottomrule
  \end{tabularx}
\end{table}

Validation Dice and IoU in Supplementary Table~\ref{tab:sup1} measure agreement with the annotated recto-surface masks, before manual mesh correction.
{}

\FloatBarrier
\begin{nolinenumbers}
\begingroup\footnotesize
\raggedbottom
\begin{xltabular}{\textwidth}{@{}p{24mm}p{30mm}X@{}}
\caption{\textbf{Training configuration of the ink detection model ablation}}
\label{tab:sup2}\\
\multicolumn{3}{@{}p{\dimexpr\textwidth-2\tabcolsep\relax}@{}}{Values pulled from the wandb run vesuvius-challenge/paper/l2\_ink5\_l5infer (rd0qz8ps), created 2026-06-15, fine-tuned from the ResNet3D-50 release r3d50\_KM\_200ep.pth, pretrained on the combined Kinetics-700 and Moments in Time datasets\cite{ref42,kataoka2020mega}.}\\[0.6em]
\toprule
Category & Parameter & Value \\
\midrule
\endfirsthead
\multicolumn{3}{@{}l}{\textit{\tablename\ \thetable\ (continued)}}\\
\toprule
Category & Parameter & Value \\
\midrule
\endhead
\bottomrule
\endfoot
\bottomrule
\endlastfoot
\multirow{6}{*}{\textbf{Architecture}}
      & Backbone & ResNet3D-50\cite{ref42} — 3D conv, BN, ReLU residual blocks \\
      & Decoder & 2D U-Net: 3 upsampling blocks of (bilinear ×2 → concat skip → 3×3 conv → BN → ReLU) \\
      & Encoder--decoder bridge & per-stage max over the z (depth) axis → 4-D feature pyramid \\
      & Convolution & nn.Conv3d (encoder), nn.Conv2d (decoder) \\
      & Normalization & nn.BatchNorm3d (encoder), nn.BatchNorm2d (decoder) \\
      & Encoder init & r3d50\_KM\_200ep.pth (Kinetics-700 and Moments in Time); conv1 weights summed across RGB → 1 grayscale channel \\
    \cmidrule(lr){2-3}
    \multirow{5}{*}{\textbf{Input}}
      & Output channels & 1 (sigmoid binary ink probability) \\
      & Patch size (D × H × W) & 62 × 256 × 256 voxels \\
      & Intensity preprocessing & layer values clipped to [0, 200] in data.py, then Normalize(mean=0, std=1) per voxel \\
      & Z-layer range & layers 1--62 from each segment's layer stack \\
      & Ignore label & none (binary supervision) \\
    \cmidrule(lr){2-3}
    \multirow{8}{*}{\makecell[tl]{\textbf{Training data}\\\textit{(ink5 row)}}}
      & Segment & l\_2 (\textit{PHerc.}~fragment) \\
      & Label file & labels/l\_2\_inklabels5.png \\
      & Train tiles & 33,061 (256×256 sub-tiles at stride 64) \\
      & Min labeled-ratio gate & label must have ≥ 1 non-zero pixel after binarization \\
      & Min bbox-fill gate & tile fully inside the segment fragment mask \\
      & Train / val split & 100 \% train; held-out segment l\_5 is used for inference only (no labels) \\
      & Inference tiles on l\_5 & 86,765 (stride 128, 4× oversample) \\
      & Inference tiles on l\_2 & 22,204 (stride 128, after fragment-mask filter) \\
    \cmidrule(lr){2-3}
    \multirow{4}{*}{\makecell[tl]{\textbf{Augmentations}\\\textit{(training only)}}}
      & Flip & HorizontalFlip(p=0.5), VerticalFlip(p=0.5) \\
      & Affine & ShiftScaleRotate(rot ±360°, shift ±0.15, scale ±0.10, p=0.75) \\
      & Blur & one of \{GaussianBlur, MotionBlur\} (p=0.4) \\
      & Cutout & CoarseDropout(max\_holes=2, max\_size=20 \% of patch, p=0.5) \\
    \cmidrule(lr){2-3}
    \multirow{15}{=}{\textbf{Optimization}}
      & Loss & 0.5 × Dice + 0.5 × SoftBCE (smooth = 0.25), against label down-interpolated to 64×64 \\
      & Optimizer & AdamW \\
      & Initial LR & $2\times10^{-5}$ \\
      & Max LR (OneCycle) & $3\times10^{-4}$ \\
      & LR schedule & OneCycleLR, pct\_start = 0.15, final\_div\_factor = $10^{2}$ \\
      & Batch size & 2 \\
      & Gradient accumulation & 4 (effective batch = 8) \\
      & Mixed precision & 16-mixed (PyTorch AMP) \\
      & Gradient clipping (L2 norm) & 1.0 \\
      & Max steps (scheduled) & 12,396 \\
      & Steps per epoch (ink5) & 4,132 \\
      & Epoch budget (ink5) & 3 \\
      & Random seed & 130697 \\
      & Determinism & cudnn.deterministic = True, cudnn.benchmark = False \\
      & Compute & GPU: 1 × NVIDIA H100 80 GB \\
    \cmidrule(lr){2-3}
    \multirow{3}{*}{\textbf{Compute}}
      & Per-run memory & ≈ 12 GB GPU, ≈ 160 GB host RAM (l\_2 + l\_5 layer stacks held in process memory) \\
      & Per-run wall time & ≈ 2 h end-to-end (load + train + l\_5 inference) \\
      & Parallelization & 5 ablation runs + 1 cross-segment baseline launched simultaneously across GPUs 0--5 \\
    \cmidrule(lr){2-3}
    \multirow{3}{*}{\textbf{Outputs}}
      & Checkpoint & checkpoints/trained/l2\_ink5\_last.ckpt (963 MB) \\
      & l\_2 prediction & predictions/l\_2/l2\_ink5\_l\_2\_pred.\{png,npy\} \\
      & l\_5 prediction & predictions/l\_5/l2\_ink5\_l\_5\_pred.\{png,npy\} \\
    \cmidrule(lr){2-3}
    \makecell[tl]{\textbf{Final training loss}\\\textit{(ink5, step 12,396)}}
      & Train total loss & 0.5299 \\
\end{xltabular}
\endgroup
\end{nolinenumbers}

The held-out segment $\ell_5$ in Supplementary Table~\ref{tab:sup2} was used for qualitative inference only; the reported final loss is the training loss on $\ell_2$.

\FloatBarrier
\begin{table}[htbp]
\centering
\footnotesize
\caption{\textbf{Canonical ink detector and inference for the five \textit{PHerc.}~139 panels in Extended Data Fig.~\ref{fig:sup1}.} Model and checkpoint metadata are from the \href{https://huggingface.co/scrollprize/ink_canonical_2um}{canonical model card}; inference and display settings are from the supplied panel metadata.}
\label{tab:139canonical}
\begin{tabularx}{\textwidth}{@{}l l X@{}}
\toprule
Category & Parameter & Value \\
\midrule
Model & Checkpoint & \texttt{r152\_3ddec\_v2\_l5\_epoch13.ckpt} \\
 & Architecture & ResNet3D-152 encoder, 3D decoder, learned depth-attention pooling and 2D output head \\
 & Recipe & \texttt{new\_canon\_autoresearch\_recipe} \\
 & Checkpoint progress & Epoch 13; global step 113,246 \\
\midrule
Input & Source & 65-layer \textit{PHerc.}~139 surface volumes; 2.399~µm pixels \\
 & Depth and normalization & Layers 63 down to 2 (62 layers); clip values at 200 and divide by 200 \\
\midrule
Inference & Tiling & 256-pixel tiles, stride 128, Hann blending \\
 & Execution & Half precision; no test-time augmentation \\
\midrule
Display & CT composite & Maximum projection of 65 layers; first and 99th percentiles mapped to gray levels 91 and 194 \\
 & Overlay & Pixelwise minimum of composite and inverted ink prediction \\
\bottomrule
\end{tabularx}
\end{table}

\FloatBarrier
\begin{nolinenumbers}
\begingroup\footnotesize
\raggedbottom
\begin{xltabular}{\textwidth}{@{}l X X@{}}
\caption{\textbf{Training configuration of the 3D DINOv2 representation model (dinovol).} Values were taken from the Weights \& Biases pre-training run model\_v2\_90k\_shift005\_jitter105 (model\_v2\_\_shift005\_jitter105, 2026-03-27), with \textit{PHerc.~Paris.}~4 finetuning from model\_v2\_90k\_shift005\_jitter105\_r342500\_paris4\_20260416 (\ldots\_\_r342500\_\_paris4\_\_20260416, 2026-04-16). Both runs are archived in project vesuvius-challenge/paper.}
\label{tab:sup3}\\
\toprule
Category & Parameter & Value \\
\midrule
\endfirsthead
\multicolumn{3}{@{}l}{\textit{\tablename\ \thetable\ (continued)}}\\
\toprule
Category & Parameter & Value \\
\midrule
\endhead
\bottomrule
\endfoot
\bottomrule
\endlastfoot
\multirow{9}{*}{Architecture}
      & Backbone & 3D Vision Transformer (DINOv2 / EVA-style, "v2" config), single-channel volumetric input \\
      & Patch embedding & nn.Conv3d, 8 × 8 × 8-voxel patches (non-overlapping) \\
      & Embedding dimension & 864 \\
      & Depth (transformer blocks) & 24 \\
      & Attention heads & 16 \\
      & Attention & Multi-head self-attention with mixed 3D rotary position embeddings (RoPE) \\
      & MLP & SwiGLU, ratio 8/3 (≈ 2.667) \\
      & Normalization & LayerNorm (pre-norm) \\
      & Register tokens & 4 \\
    \cmidrule(lr){2-3}
    \multirow{2}{*}{RoPE settings}
      & base 100; per-axis ("separate") normalization & train-time coord. aug. shift 0.05 / jitter 1.05 / rescale 2 \\
      & Stochastic depth (drop-path) & 0.2 \\
    \cmidrule(lr){2-3}
    \multirow{7}{*}{Input}
      & Input channels & 1 (grayscale CT) \\
      & Global crops (Z × Y × X) & 128 × 128 × 128 voxels, × 2 \\
      & Local crops & 64 × 64 × 64 voxels, × 8 (scale range 0.02--0.25) \\
      & Source sampling region & 256 × 256 × 256 voxels \\
      & Intensity normalization & Per-volume robust scaling (median / IQR) \\
      & Voxel size & ≈ 2.4 µm isotropic \\
    \cmidrule(lr){2-3}
    \multirow{4}{*}{\makecell[tl]{Self-supervised\\objective}}
      & Losses & DINO (CLS-token) + iBOT (masked-patch) + KoLeo regularizer (Gram term present, weight 2, inactive) \\
      & Centering & Sinkhorn--Knopp \\
      & Teacher temperature & 0.04 → 0.07 (warmup 25,000 steps) \\
      & Teacher EMA momentum & 0.994 → 1.0 \\
    \cmidrule(lr){2-3}
    \multirow{3}{*}{Training data}
      & Pretraining volumes (10 scrolls, ESRF ≈ 2.4 µm) & \textit{PHerc.}~9B, 500P2, 814, 1299, 343P, 332, 139, MAN5, 1451, MANBp \\
      & Validation volumes & \textit{PHerc.}~332, \textit{PHerc.}~139 \\
      & \textit{PHerc.~Paris.}~4 finetuning volume & \textit{PHerc.~Paris.}~4 (20260411\ldots, 2.400 µm) — added at finetuning \\
    \cmidrule(lr){2-3}
    \multirow{9}{*}{Optimization}
      & Optimizer & AdamW \\
      & Learning rate & $2.5\times10^{-5}$ (cosine; warmup 35,000 steps → min $1\times10^{-6}$) \\
      & Patch-embedding LR multiplier & 0.2 \\
      & Weight decay & 0.04 → 0.2 (cosine) \\
      & Gradient clipping (L2 norm) & 5 \\
      & Batch size & 3 (per device) \\
      & Mixed precision & AMP \\
      & Max iterations (scheduled) & 1,000,000 \\
      & Checkpoint / validation cadence & every 2,500 / 500 steps \\
    \cmidrule(lr){2-3}
    \multirow{3}{*}{\makecell[tl]{Pretraining\\end-state}}
      & Final logged step & 342,558 (end-state: stopped; checkpoint\_step\_342500 used downstream) \\
      & Learning rate at final step & $1.95\times10^{-5}$ \\
      & Train / Val loss total & 14.566 / 16.008 \\
    \cmidrule(lr){2-3}
    \multirow{2}{*}{\makecell[tl]{\textit{PHerc.~Paris.}~4\\finetuning}}
      & Steps & resumed at 342,500 → final 353,686 (≈ 11,200 steps) \\
      & Validation loss (total) & 15.076 \\
\end{xltabular}
\endgroup
\end{nolinenumbers}

Validation losses in Supplementary Table~\ref{tab:sup3} monitor the self-supervised objective on randomly sampled crops from volumes also used in training. Pretraining validation used \textit{PHerc.}~332 and \textit{PHerc.}~139; during \textit{PHerc.~Paris.}~4 fine-tuning, it also included \textit{PHerc.~Paris.}~4.
{}

\FloatBarrier
\begin{nolinenumbers}
\begingroup\footnotesize
\raggedbottom
\begin{xltabular}{\textwidth}{@{}l X X@{}}
\caption{\textbf{Training configuration of the \textit{PHerc.~Paris.}~4 volumetric 3D ink-detection models.} The four models share the architecture and optimization settings listed below and differ only in their supervision. Configurations were taken from the corresponding Weights \& Biases runs, archived in project vesuvius-challenge/paper: feasible-voice-35 (76ezvyks, 2026-04-17), morning-rain-52 (6az0505g, 2026-04-24), divine-butterfly-57 (af3pga3d, 2026-04-24) and chocolate-wildflower-62 (4b07qv8p, 2026-04-27).}
\label{tab:sup4}\\
\toprule
Category & Parameter & Value \\
\midrule
\endfirsthead
\multicolumn{3}{@{}l}{\textit{\tablename\ \thetable\ (continued)}}\\
\toprule
Category & Parameter & Value \\
\midrule
\endhead
\bottomrule
\endfoot
\bottomrule
\endlastfoot
\multirow{8}{*}{Architecture}
      & Backbone & nnU-Net-style 3D residual-encoder U-Net (vesuvius NetworkFromConfig, auto-configured) \\
      & Convolution & nn.Conv3d \\
      & Normalization & nn.InstanceNorm3d \\
      & Encoder stages / feature widths & 7 stages — 32, 64, 128, 256, 320, 320, 320 \\
      & Residual blocks per stage & 1, 3, 4, 6, 6, 6, 6 \\
      & Decoder & Shared decoder, skip-concatenation; single ink head (1 × 1 × 1 conv) \\
      & Parameters & $1.42\times10^{8}$ (≈ 142 M) \\
      & Output channels & 1 (ink; sigmoid + BCE) \\
    \cmidrule(lr){2-3}
    \multirow{5}{*}{Input}
      & Patch size (Z × Y × X) & 256 × 256 × 256 voxels \\
      & Input channels & 1 \\
      & Intensity normalization & Per-volume percentile min--max \\
      & Mode & full-3D (surface annotations placed ± 3 voxels about the reference surface) \\
      & Min labeled-coverage gate & 0.02 \\
    \cmidrule(lr){2-3}
    \multirow{3}{*}{Training data}
      & Teacher supervision (run 76ezvyks) & Surface-conditioned 2D-ink dataset projected into 3D — 8 scrolls: \textit{PHerc.~Paris.}~4, 139, 500P2, 814, 841, 1667, MAN5, 9B \\
      & Guidance / distillation volume (6az0505g, af3pga3d, 4b07qv8p) & \textit{PHerc.~Paris.}~4 \\
      & Seed & 27 \\
    \cmidrule(lr){2-3}
    \multirow{8}{*}{Optimization}
      & Loss & LabelSmoothedDCAndBCELoss — soft Dice + binary cross-entropy on the 1-channel ink output (weight\_dice = weight\_ce = 1; Dice \& BCE label smoothing 0.1; ignore-label enabled) \\
      & Optimizer & SGD, Nesterov momentum 0.99, weight decay $3\times10^{-5}$ \\
      & Learning-rate schedule & Cosine with warmup (diffusers\_cosine\_warmup); base LR $1\times10^{-2}$; 5,000-step warmup \\
      & Batch size & 2 \\
      & Mixed precision & bf16 \\
      & Weight EMA & decay 0.9995 (from step 1,000); used for validation and as the frozen guidance teacher \\
      & Max iterations (scheduled) & 250,000 \\
    \cmidrule(lr){2-3}
    \multirow{5}{*}{\makecell[tl]{DINO-guided\\rounds}}
      & DINO model & \textit{PHerc.~Paris.}~4-finetuned dinovol (checkpoint\_step\_352500); 128³ sliding windows (stride 128) → 16³ patch tokens; Gaussian blend σ = 4; per-chunk min--max \\
      & Reference ink embedding & Mean of 256 expert-clicked 864-D DINO patch tokens, L2-normalized \\
      & Threshold τ & 0.5 (guided rounds) \\
      & Background mask & input intensity ≥ 50 (added from round 2 onward) \\
      & Threshold τ & 0.5 \\
    \cmidrule(lr){2-3}
    \multirow{4}{*}{\makecell[tl]{Pipeline stages\\\& run end-state}}
      & Stage 1 · teacher (76ezvyks) & 8-scroll full supervision; final step 68,000; val balanced-acc. 0.664; val loss 0.426 (EMA 0.396) \\
      & Stage 2 · DINO-guided student (6az0505g) & init from teacher; labels = σ(teacher)·minmax(cosine→ink) > 0.5; final step 62,950; val balanced-acc. 0.715; val loss 0.317 (EMA 0.131) \\
      & Stage 3 · + background mask (af3pga3d) & init from stage 2 (also its frozen teacher); mask intensity < 50; final step 77,000; val balanced-acc. 0.576; val loss 1.504 (EMA 1.487) \\
      & Stage 4 · self-distillation (4b07qv8p) & init from stage 3; pseudo-labels from ensemble of stage-3 checkpoints (steps 77k + 64k) + TTA, DINO dropped, full supervision; final step 78,450; val balanced-acc. 0.603; val loss 0.352 (EMA 0.303) \\
\end{xltabular}
\endgroup
\end{nolinenumbers}

{}